 \documentclass[preprint2]{aastex}

\shorttitle{Analysis of optical \ion{Fe}{2} emission in a sample of AGN spectra}
\shortauthors{Kova\v cevi\' c et al.}

\begin{document}


\title{ Analysis of optical \ion{Fe}{2} emission in a sample of AGN spectra}


\author{Jelena Kova\v cevi\' c\altaffilmark{1,2}, Luka \v C. Popovi\' c\altaffilmark{1,2} and Milan S. Dimitrijevi\' c\altaffilmark{1,2,3}}

\affil{\altaffilmark{1}Astronomical  Observatory,  Volgina  7, 11060  Belgrade, Serbia}
\affil{\altaffilmark{2}Isaac Newton Institute of Chile, Yugoslavia Branch}
\affil{\altaffilmark{3}Laboratoire d'\'{E}tude du Rayonnement et de la Mati\`{e}re
en Astrophysique, UMR CNRS 8112, Observatoire de Paris-Meudon,
92195 Meudon, France}
\email{jkovacevic@aob.bg.ac.rs}



\begin{abstract}

We present a study of optical \ion{Fe}{2} emission in 302 AGNs selected from the SDSS. We group the strongest \ion{Fe}{2} multiplets into three groups according to the lower term of the transition (b${\ }^4F$, a${\ }^6S$ and a${\ }^4G$ terms). These correspond approximately to the blue, central, and red part respectively of the ``iron shelf'' around H$\beta$. We calculate an \ion{Fe}{2} template which takes into account transitions into these three terms and an additional group of lines, based on a reconstruction of the spectrum of I Zw 1. This \ion{Fe}{2} template gives a more precise fit of the \ion{Fe}{2} lines in broad-line AGNs than other templates. We extract \ion{Fe}{2}, H$\alpha$, H$\beta$, [\ion{O}{3}] and [\ion{N}{2}] emission parameters and investigate correlations between them. We find that \ion{Fe}{2} lines probably originate in an Intermediate Line Region. We notice that the blue, red, and central parts of the iron shelf have different relative intensities in different objects. Their ratios depend on continuum luminosity, FWHM H$\beta$, the velocity shift of \ion{Fe}{2}, and the $\mathrm{H\alpha}$/$\mathrm{H\beta}$ flux ratio. We examine the dependence of the well-known anti-correlation between the equivalent widths of \ion{Fe}{2} and [\ion{O}{3}] on continuum luminosity. We find that there is a Baldwin effect for [\ion{O}{3}] but an inverse Baldwin effect for the \ion{Fe}{2} emission. The [\ion{O}{3}]/\ion{Fe}{2} ratio thus decreases with L$_{\lambda5100}$. Since the ratio is a major component of the Boroson and Green eigenvector 1, this implies a connection between the Baldwin effect and eigenvector 1, and could be connected with AGN evolution. We find that spectra are different for H$\beta$ FWHMs greater and less than $\sim$3000 $\mathrm{kms^{-1}}$, and that there are different correlation coefficients between the parameters.

\end{abstract}



\section{Introduction}
Optical \ion{Fe}{2} ($\lambda\lambda$ 4400-5400 \AA) emission is one of the most interesting features in AGN spectra. The emission arises from numerous transitions of the complex \ion{Fe}{2} ion. \ion{Fe}{2} emission is seen in almost all type-1 AGN spectra and it is especially strong in narrow-line Seyfert 1s (NLS1s). It can also appear in the polarized flux of type-2 AGNs when a hidden BLR can be seen. Origin of the optical \ion{Fe}{2} lines, the mechanisms of their excitation and location of the \ion{Fe}{2} emission region in AGN, are still open questions. There are also many correlations between \ion{Fe}{2} emission and other AGN properties which need a physical explanation.

Some of the problems connected with \ion{Fe}{2} emission are:

 1.  There have been suggestions that \ion{Fe}{2} emission cannot be explained with standard photoionization models \citep[see][and references therein]{b60,bb}. To explain the \ion{Fe}{2} emission, additional mechanisms of excitation have been proposed:  a) continuum fluorescence \citep{b1,b2},  b) collisional excitation \citep{b3,b4,b5,b6}, c) self-fluorescence among \ion{Fe}{2} transitions \citep{b7},  d) fluorescent excitation by the Ly$\alpha$ and Ly$\beta$ lines \citep{b8,b300}.

2. There are uncertainties about where the \ion{Fe}{2} emission line region is located in an AGN. A few proposed solutions are that the \ion{Fe}{2} lines arise: a) in the same emission region as the broad H$\beta$ line \citep{b9}, b) in the accretion disk near the central black hole, which produces the double-peaked broad Balmer emission lines  \citep{bb, b10}, c) in a region which can be heated by shocks or from an inflow, located between BLR and NLR -- the so-called the Intermediate Line Region \citep{b001,b31,b66,b222,b51,b151} and d) and in the shielded, neutral, outer region of a flattened BLR \citep{baa,ba}.

3. It has long been established that the \ion{Fe}{2} emission is  correlated with the radio, X and IR continuum. \ion{Fe}{2} lines are stronger in spectra of Radio Quiet (RQ) AGNs, than in Radio Loud (RL) objects \citep{b11,b12}. But, if we consider RL AGNs, \ion{Fe}{2} emission is stronger in Core Dominant (CD) AGNs than in Lobe Dominant (LD) AGNs \citep{b13,b5}.  The relative strength of the \ion{Fe}{2} lines correlates with the soft X-ray slope, but anti-correlates with X-ray luminosity \citep{b14,bc,b15}. Strong optical \ion{Fe}{2} emission is usually associated with strong IR luminosity \citep{b17,b19}, but it is anti-correlated with IR color index $\alpha$(60,25) \citep{be}. 

4.  Many correlations have been observed between \ion{Fe}{2} and other emission lines. \citet{b9} give a number of correlations between the  equivalent width (EW) of \ion{Fe}{2} and properties of the [\ion{O}{3}] and H$\beta$ lines. The most interesting are the anti-correlations of EW \ion{Fe}{2} vs. EW [\ion{O}{3}]/EW H$\beta$, EW \ion{Fe}{2} vs. peak [\ion{O}{3}], EW \ion{Fe}{2}/EW H$\beta$ vs. peak [\ion{O}{3}], and the correlation of EW \ion{Fe}{2}/EW H$\beta$ vs. H$\beta$ asymmetry. An anti-correlation between EW \ion{Fe}{2} and FWHM of H$\beta$ has been also found \citep{b20,b21,b9}.  These correlations dominate the first eigenvector of the principal component analysis of  \citet{b9} (hereinafter EV1). Correlations between EV1 and some properties of C IV $\lambda1549$ \AA \ lines have also been pointed out \citep{b22}. With increasing continuum luminosity, the EW of the C IV lines decreases, \citep[the so-called ``Baldwin effect'',][]{b35,b23}, while the blueshift of the C IV line and the EW of \ion{Fe}{2} increase.  Recently, \citet{bf} have found the Baldwin effect to be related to EV1. \ion{Fe}{2} lines are usually strong in AGN spectra with low-ionization Broad Absorption Lines (BALs) \citep{b24,b18}.

Searching for a physical cause of these correlations may help in understanding of the origin of \ion{Fe}{2} emission. Some of physical properties which may influence these correlations are: Eddington ratio (L/$\mathrm{L_{Edd}}$) \citep{b9,b22}, black hole mass ($M_{BH}$) and inclination angle \citep{b25,b26,b27}.

In addition to these properties, \ion{Fe}{2} emission strength may vary with evolution of AGNs. For example, \citet{b29} have explained some properties of AGN (as BALs, \ion{Fe}{2} strength, radio properties, emission line width, luminosity of narrow lines, etc.) by an ``evolutive unification model''. In this model, accretion arises from the interaction between nuclear starbursts and the supermassive black hole.  This would mean that not only the orientation, but also the evolutionary state of an AGN, influence its spectral properties. Thus, young AGN have strong \ion{Fe}{2}, BALs, weak radio emission, the NLR is compact and faint, and broad lines are relatively narrow. In contrast with this, old AGNs have weak \ion{Fe}{2}, no BALs, strong radio emission with extended radio lobes, the NLR is extended and bright, and the broad lines have greater velocity widths.

 In this paper, we investigate the \ion{Fe}{2} emitting region by analyzing the correlations between the optical \ion{Fe}{2} lines and the other emission lines in a sample of 302 AGN from the SDSS. To do this we construct an \ion{Fe}{2} template. The strongest \ion{Fe}{2} multiplets within the $\lambda\lambda$4400-5500 \AA \ range are sorted into three groups, according to the lower terms of the transitions: $3d^6({\ }^3F2)4s{\ }^4F$, $3d^54s^2{\ }^6S$ and $3d^6({\ }^3G)4s{\ }^4G$ (which we will hereinafter refer to as the $F$, $S$ and $G$ groups of lines). The $F$ group mainly contains the lines from the \ion{Fe}{2} multiplets 37 and 38 and describes the blue part of the \ion{Fe}{2} shelf relative to H$\beta$. The $S$ group of lines describes the part of \ion{Fe}{2} emission under H$\beta$ and [\ion{O}{3}] and contains lines from multiples 41, 42 and 43 \ion{Fe}{2}. Finally, the $G$ group contains lines from multiplets 48 and 49, and describes the red part of the \ion{Fe}{2} shelf. A simplified Grotrian diagram of these transitions is shown in Figure \ref{1}. We analyze separately their relationships with other lines in AGN spectra (H$\beta$, [\ion{O}{3}], H$\alpha$, [\ion{N}{2}]). In this way, we try to connect details of the \ion{Fe}{2} emission with physical properties of the AGN emission regions in which these lines arise. Each of the \ion{Fe}{2} multiplet groups has specific characteristics, which may be reflected in a different percent of correlations with other lines, and give us more information about the complex \ion{Fe}{2} emission region.

The paper is organized as follows: In Sect. 2 we describe the procedure of sample selection and details of analysis; our results are presented in Sect. 3; in Sect. 4 we give discussion of obtained results, and finally in Sect. 5 we outline our conclusions.

\section{The sample and analysis}

\subsection{The AGN sample}

Spectra for our data sample are taken from the the $7^{th}$ data release  \citep{bg} of the Sloan Digital Sky Survey (SDSS).  For the purposes of the work, we chose AGNs with the following characteristics:

\begin{enumerate}
 \item high signal to noise ratio ($\mathrm{S/N}>20$),
\item good pixel quality (profiles are not affected by distortions due to bad pixels on the sensors, the presence of strong foreground or background sources),
 \item high redshift confidence (zConf$>$0.95) and with $z\leq 0.7$ in order to cover the optical \ion{Fe}{2} lines around H$\beta$ and [\ion{O}{3}] lines,
\item  negligible contribution from the stellar component.  We controlled for this by having the EWs of typical absorption lines be less then 1 \AA \ (EW CaK 3934 \AA, Mg 5177 \AA \ and H$\delta$ 4102 \AA $>$-1).\footnote { For some objects we were able to find estimates of the  stellar component in the literature and to confirm that the host galaxy fraction is indeed small in our sample (see Appendix A).}  Because of this it was not necessary to  remove a host galaxy starlight contribution.
\item presence of the narrow [\ion{O}{3}] and the broad H$\beta$ component (FWHM $\mathrm{H\beta>1000\ km s^{-1}}$).
\end{enumerate}

\begin{figure*}
\includegraphics[width=0.40\textwidth]{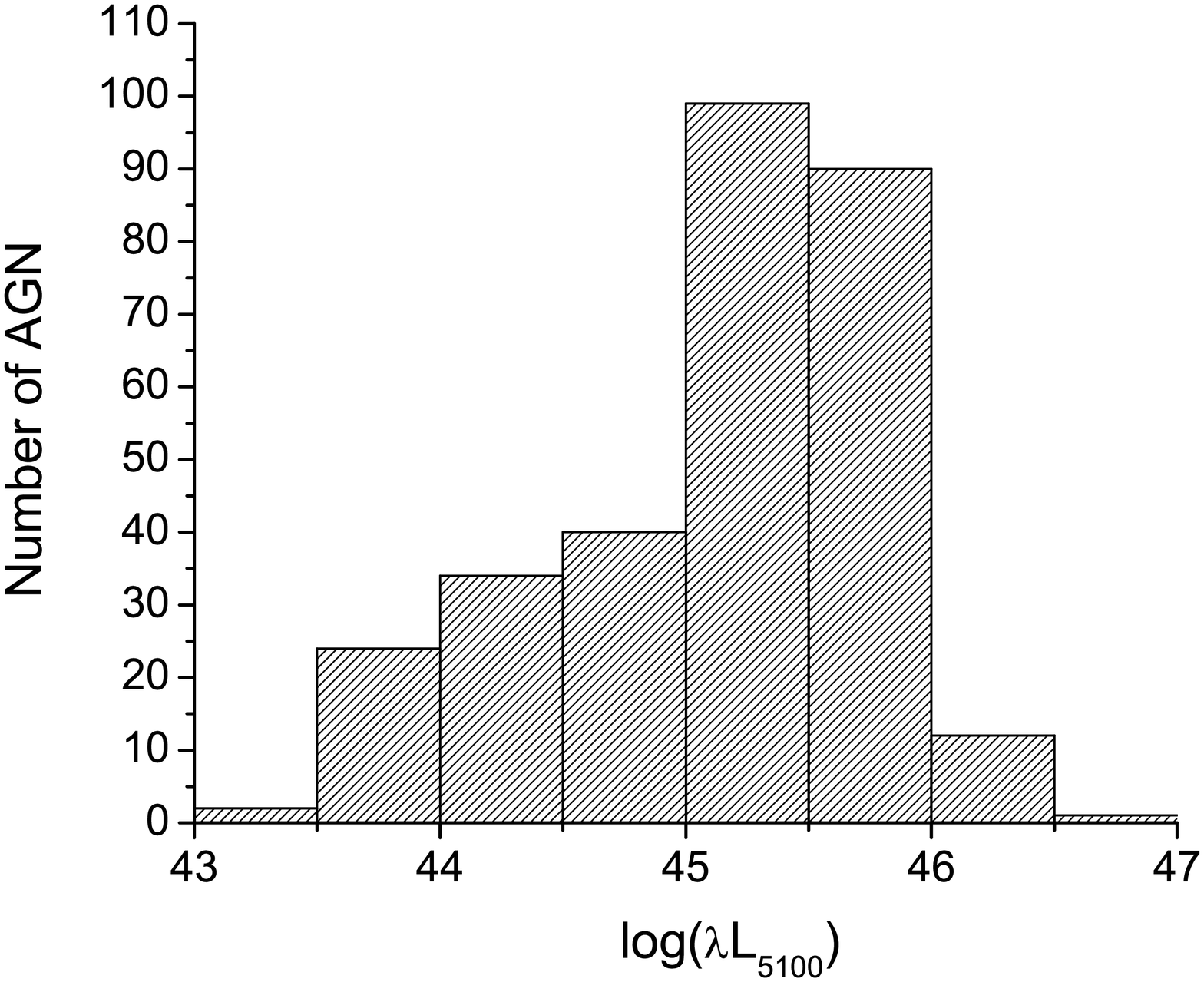}
\includegraphics[width=0.37\textwidth]{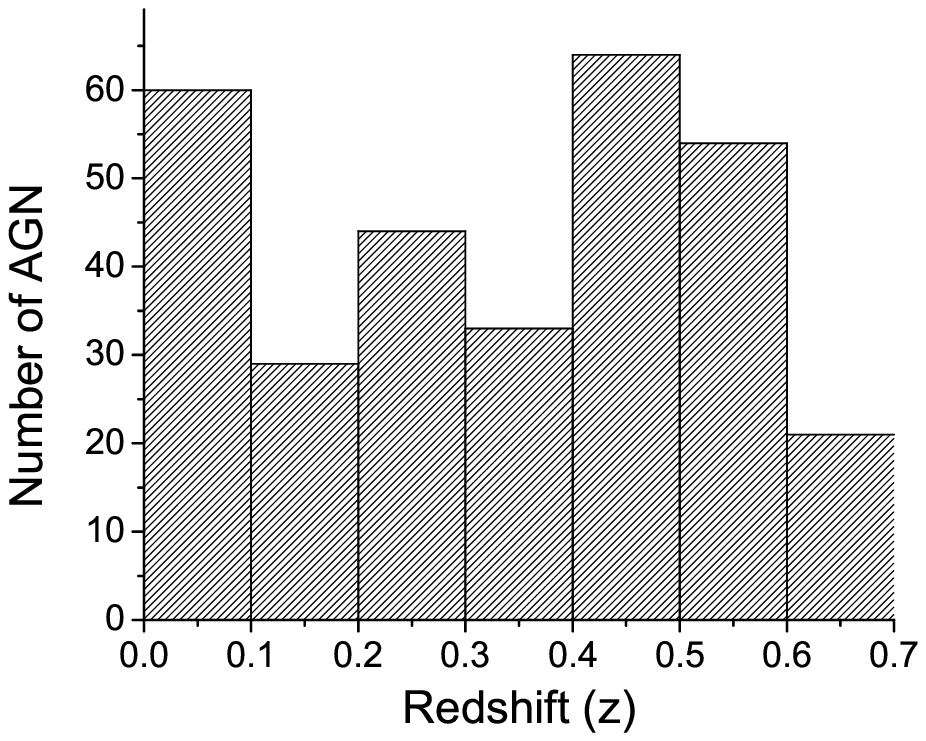}
\caption{ Distribution of continuum luminosity (left) and cosmological redshift (right) in the AGN sample.}
\label{3}
\end{figure*}

We found 497 AGNs  using the mentioned criteria.  However, on inspection of the spectra we found that in some of them the \ion{Fe}{2} line emission is within the level of noise and could not be properly fitted, so we excluded these objects from further analysis (see Appendix A).  As result, our final sample contains 302 spectra, from which 137 have all Balmer lines, and the rest are without the H$\alpha$ line (due to cosmological redshifts). The sample distribution by continuum luminosity ($\lambda$ 5100 \AA) and by cosmological redshift is given in Figure \ref{3}. Corrections for Galactic extinction were made using an empirical selective extinction function computed for each spectrum on the basis of Galactic extinction coefficients given by \citet{b34} and available
from the NASA/IPAC Extragalactic Database\footnote{http://nedwww.ipac.caltech.edu/}. Finally, all spectra were de-redshifted.

\begin{figure*}

\includegraphics[width=0.47\textwidth,angle=0]{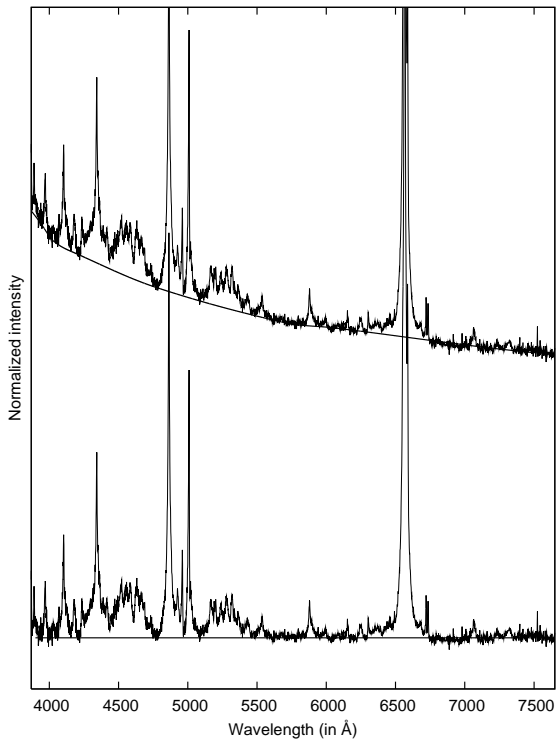}
\caption{An example of removing the continuum from the spectrum of SDSS J$075101.42+29174000.00$; The level of the continuum is determined by interpolating between the chosen points at wavelengths without emission lines (top). The spectrum without continuum emission is shown at the bottom; }
\label{2}
\end{figure*}

To subtract the continuum, we used the DIPSO software, finding the continuum level by using continuum windows given in the paper of \citet{b301} (see Figure \ref{2}). The used continuum windows are: 3010-3040 \AA, 3240-3270 \AA, 3790-3810 \AA, 4210-4230 \AA, 5080-5100 \AA, 5600-5630 \AA \ and 5970-6000 \AA.  We used the same procedure to subtract the continuum in all objects.  Continuum subtraction may cause systemic errors in the estimation of line parameters and continuum luminosity.  We estimated that the error bars are smaller than 5\%, especially within the H$\alpha$+[N II] region and in the red part of the ``iron shelf''.

We considered two spectral ranges: $\lambda\lambda$ 4400-5500 \AA \ and $\lambda\lambda$ 6400-6800 \AA. In the first range, dominant lines are the numerous \ion{Fe}{2} lines, [\ion{O}{3}] $\lambda\lambda$4959, 5007 \AA, H$\beta$ and He II $\lambda$4686 \AA, and in the second range, H$\alpha$ and [\ion{N}{2}] $\lambda\lambda$6548, 6583 \AA.

 To investigate correlations between the \ion{Fe}{2} multiplets and other lines in spectra, we fit all considered lines with Gaussians. The optical \ion{Fe}{2} lines were fitted with a template calculated as described in the next section.

\subsection{The \ion{Fe}{2} ($\lambda\lambda$ 4400-5500 \AA) template}

A number of authors have created an \ion{Fe}{2} template in the UV and optical range \citep[see][and references therein]{b50}. \citet{b9} applied empirical template by removing all lines which are not \ion{Fe}{2}, from the spectrum of I Zw 1. Similarly, \citet{b303} constructed an \ion{Fe}{2} template by identifying systems of broad and of narrow \ion{Fe}{2} lines in the I Zw 1 spectrum, and measuring their relative intensities. \citet{b302} improve on that template by using two parameters of intensity - one for the broad, one for the narrow \ion{Fe}{2} lines. All these empirical templates are defined by the line width and the line strength, implying that relative strengths of the lines in the \ion{Fe}{2} multiplets are the same in all objects.

 In theoretical modeling, significant effort has been made to calculate the iron emission by including a large number of \ion{Fe}{2} atomic levels, going to high energy \citep{b300,b32,bb,b304}.
\citet{b304} calculated \ion{Fe}{2} emission using CLOUDY code and 830 level \ion{Fe}{2} model. In this model, energies go up to 14.6 eV, producing 344,035 atomic transitions. For their calculations they used solar abundances for a range of physical conditions such as the flux of ionizing photons [$\mathrm{\Phi_H}$], hydrogen density [$\mathrm{n_H}$], and microturbulence [$\xi$].

    Using existing \ion{Fe}{2} templates, we found that empirical and theoretical
models can generally fit NLSy1 \ion{Fe}{2} lines well, but that in some cases of spectra with broader H$\beta$ lines,
the existing models did not provide as good fit.

    One of the problems in the analysis of \ion{Fe}{2} emission is that it consists of
numerous overlapping lines. This makes the identification
and determination of relative intensities very difficult. Therefore, the list of \ion{Fe}{2} lines used for the fit of the \ion{Fe}{2} emission, as well as their relative intensities,
are different in different models \citep{b303,b304}. Also, there is significant disagreement in values of oscillator strengths in different atomic data sources \citep{b100, b105, b101}.

    To investigate the \ion{Fe}{2} emission, we made an \ion{Fe}{2} template taking
into account following: (a)  majority of multiplets dominant in the optical part ($\lambda\lambda$ 4400-5500 \AA), whose lines can be clearly identified in AGN spectra, have one of the three specific lower terms of their transitions: $F$, $S$ or $G$, and (b) beside these lines there are also lines whose origin is not well known but which presumably originate from higher levels.

\begin{figure*}
\includegraphics[width=0.47\textwidth]{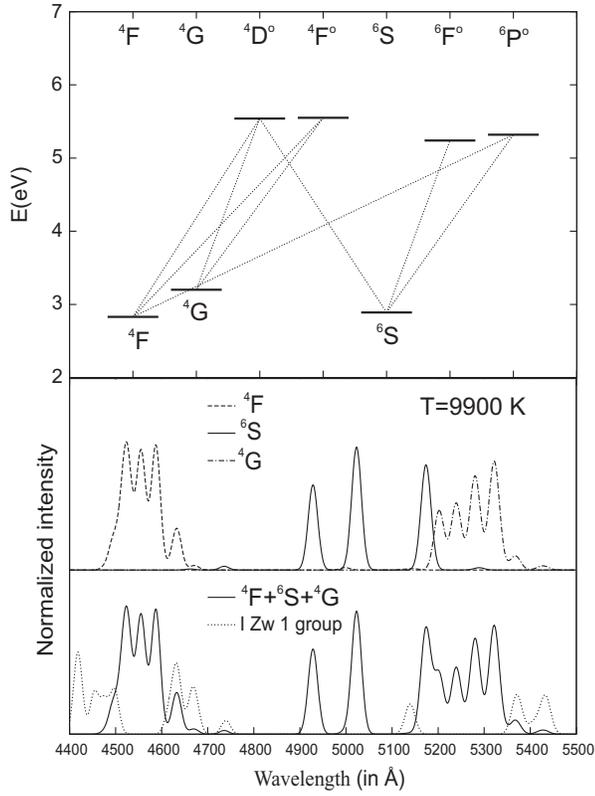}
\caption{Grotrian diagram showing the strongest \ion{Fe}{2} transitions in the $\lambda\lambda$ 4400-5500 \AA \ region (top). Lines are separated into three groups according to the lower level of transition (middle): $F$ (dashed line), $S$ (solid line) and $G$ (dash--dotted line). Bottom: the lines from the three line groups (solid line) and lines measured from I Zw 1, represented with dots.}
\label{1}
\end{figure*}

 We constructed an \ion{Fe}{2}  template consisting of 50 \ion{Fe}{2} emission lines, identified as the strongest within $\lambda\lambda$ 4400-5500 \AA \ range. 35 of them are sorted into three line groups according to the lower term of their transition ($F$, $S$ and $G$). The $F$ group consists of 19 lines (mainly multiplets 37 and 38) and dominates in the blue shelf of the iron template (4400-4700 \AA). The $S$ group consists of 5 lines (multiplets 41, 42 and 43) and describes the \ion{Fe}{2} emission covering the [\ion{O}{3}] and H$\beta$ region of the spectrum and some of  emission from the red \ion{Fe}{2} bump (5150-5400 \AA), and the $G$ lines (11 lines from multiplets 48 and 49) dominate in the red bump (5150-5400 \AA) (see Figure \ref{1}).

We assume that the profiles of each of lines can be represented by a Gaussian, described by width ($W$), shift ($d$)\footnote{Here we used $W=\frac{W_D}{\lambda_0}$, $d=\frac{\Delta\lambda}{\lambda_0}$, where $W_D=\sigma\sqrt{2}$ is the Doppler width, $\Delta\lambda$ is the shift and $\sigma$ is the velocity dispersion} and intensity ($I$). Since all \ion{Fe}{2} lines from the template probably originate in the same region, with the same kinematical properties, values of $d$ and $W$ are assumed to
be the same for all \ion{Fe}{2} lines in the case of one AGN.

  Since the population of the lower term is influenced by transition probabilities and excitation temperature, we assumed that the relative intensities between lines within a line group ($F$, $S$ or $G$) can be obtained approximately from:
 \begin{equation}
 \frac{I_1}{I_2}={(\frac{\lambda_2}{\lambda_1})}^3\frac{f_1}{f_2}\cdot\frac{g_1}{g_2}\cdot e^{-(E_1-E_2)/kT}
\end{equation}
where $I_1$ and $I_2$ are the intensities of lines with the same lower term, $\lambda_1$ and $\lambda_2$ are the wavelengths of the transition, $g_1$ and $g_2$ are the statistical weights for the upper energy levels, $f_1$ and $f_2$ are the oscillator strengths, $E_1$ and $E_2$ are the energies of the upper levels of transitions, $k$ is the Boltzman constant, and $T$ is the excitation temperature.  Details can be found in Appendix C.  The excitation temperature is the same for all transitions in the case of partial LTE \citep[see][]{b611}, but as it is shown in Appendix C, Eq (1) may be used also in some non-LTE cases.

Lines from the three above mentioned groups explain about 75\% of the \ion{Fe}{2} emission in the observed range (4400-5500 \AA), but about 25\% of  \ion{Fe}{2} emission cannot be explained with permitted lines for which the excitation energies are close to the lines of the three groups. The missing \ion{Fe}{2} emission is around $\sim$4450 \AA, $\sim$4630 \AA, $\sim$5130 \AA \ and $\sim$5370 \AA.

There are some indications that fluorescence processes may have a role in producing some \ion{Fe}{2} lines \citep{b300,b305}. Process like self-fluorescence, continuum-fluorescence or Ly$\alpha$ and Ly$\beta$ pumping could supply enough energy to excite the \ion{Fe}{2} lines with high energy of excitation, which could be one of the explanation for emission within these wavelength regions. To complete the template for approximately of 25\% missing \ion{Fe}{2} flux, we selected 15 lines from the Kurutz database\footnote{http://www.pmp.uni-hannover.de/cgi-bin/ssi/test/kurucz/sekur.html} with wavelengths close to those of the extra emission, upper level excitation energies of up to $\sim$13 eV, and strong oscillator strengths. We measured their relative intensities in I Zw 1 which has a  well-studied spectrum \citep{b303} spectrum with strong and narrow \ion{Fe}{2} lines. Relative intensities of these 15 lines were obtained by making the best fit of the I Zw 1 spectrum with the \ion{Fe}{2} lines from the $F$, $S$, and $G$ line groups. The extra lines are represented in Figure \ref{1} (bottom), with a dotted line.

Our template of \ion{Fe}{2} is described by 7 free fitting parameters: width, shift, four parameters of intensity (for the $F$, $S$ and $G$ line groups and for the lines with relative intensities obtained from I Zw 1). The seventh parameter is the excitation temperature included in the calculation of relative intensities within $F$, $S$ and $G$ line groups.

\begin{figure*}

\includegraphics[width=0.45\textwidth]{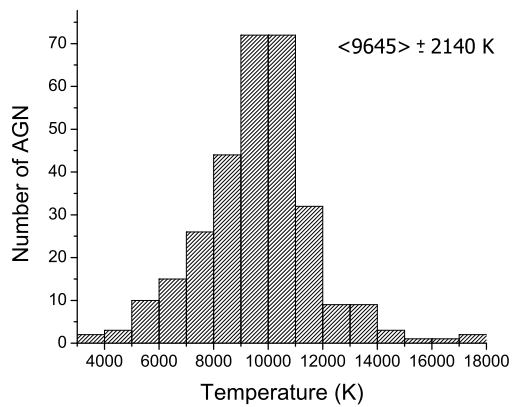}
\caption{The distribution of the excitation temperatures for the sample of 302 AGNs. Temperatures are obtained by fits using Eq. (1).  The mean value and dispersion are shown in the Figure.}
\label{4}
\end{figure*}

 We found that for the majority of objects, the excitation temperature obtained from our fit is within the range: 9000 K -- 11000 K (see Figure \ref{4}), which agrees well with theoretical predictions \citep{b60}. However, as it could be seen in Eq. (1), our fit is not very sensitive to the temperature, especially for T$>$8000 K. Eq. (1) is also very approximate so estimated temperatures should be treated with caution.

To estimate the error in the excitation temperature, we found our best fit for a number of objects.  We then changed only the excitation temperature while keeping the other fit parameters fixed.  We thus found $\chi^2_{max}$ for which the fit is still reasonable. After that, we constructed $\chi^2$ vs. $T$ and measured the dispersion in temperature at 0.1$\cdot$($\chi^2_{max}$-$\chi^2_{min}$). We obtained error-bars that were mostly within the range  $\pm25\%$, with tendency for higher temperature to have a greater uncertainty. In further analysis we will therefore not consider the temperatures obtained from the fit.

The wavelengths of the 50 template lines are presented in Tables \ref{tbl-1} and \ref{tbl-2}.
For the 35 lines from the three line groups we give multiplet names, configurations of the transitions generating those lines, oscillator strengths and the relative intensities.  The intensities have been calculated using the Eq. (1) for excitation temperatures $T$ = 5000, 10000 and 15000 K (Table \ref{tbl-1}). For the 15 lines for which relative intensities were measured in I Zw 1, we give wavelengths, configurations and oscillator strengths taken from Kurucz database, as well as their measured relative intensities (Table \ref{tbl-2}).

 The purpose of dividing the \ion{Fe}{2} emission into four groups is to investigate the correlations between the iron lines which arise from different multiplets and some spectral properties, and to compare it with results obtained for the total \ion{Fe}{2} within the observed range. Also, a template which has more parameters may give a better fit, since relative intensities among \ion{Fe}{2} lines can be different even in the similar AGN spectra, as is, for example, the case for spectra of I Zw 1 and Mrk 42 \citep[as e.g.][]{bo1}. 

\begin{figure*}
\includegraphics[width=0.65\textwidth]{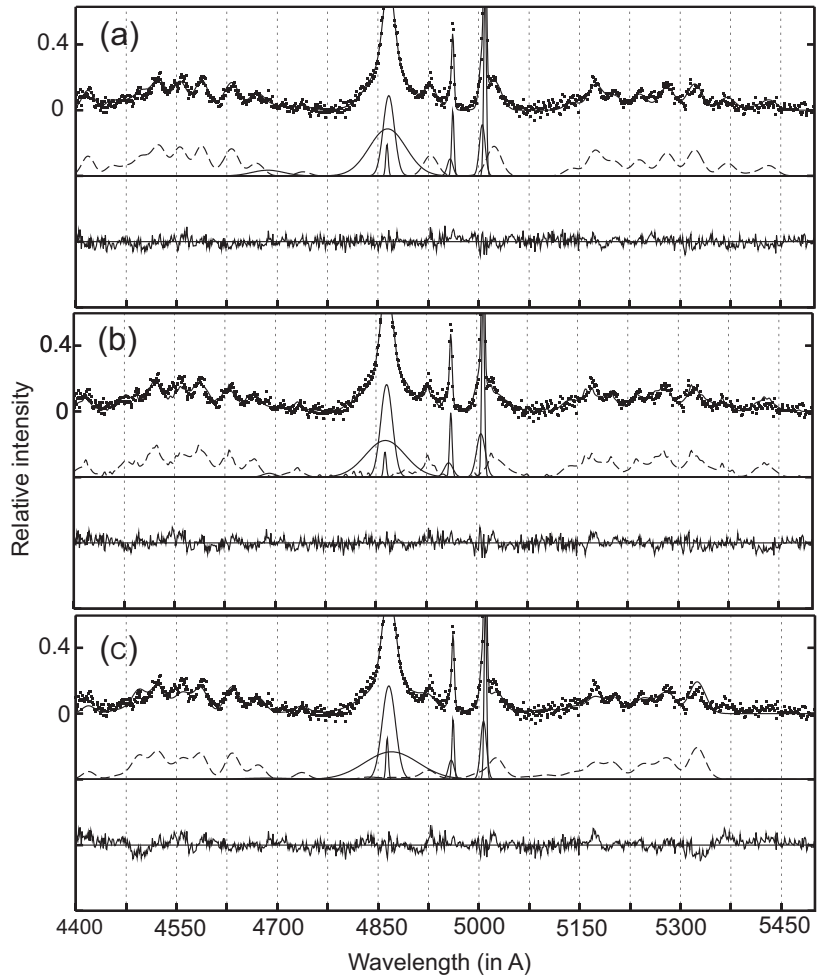}
\caption{Examples of fits to SDSS J020039.15-084554.9: with our template (a), with the empirical template of \citet{b302} (b),  and with the theoretical template of \citet{b304} (c). Since this AGN has iron emission equally strong in blue and red iron bump (as I Zw 1), all three models fit the observed lines well.}
\label{5}
\end{figure*}

\begin{figure*}
\includegraphics[width=0.65\textwidth]{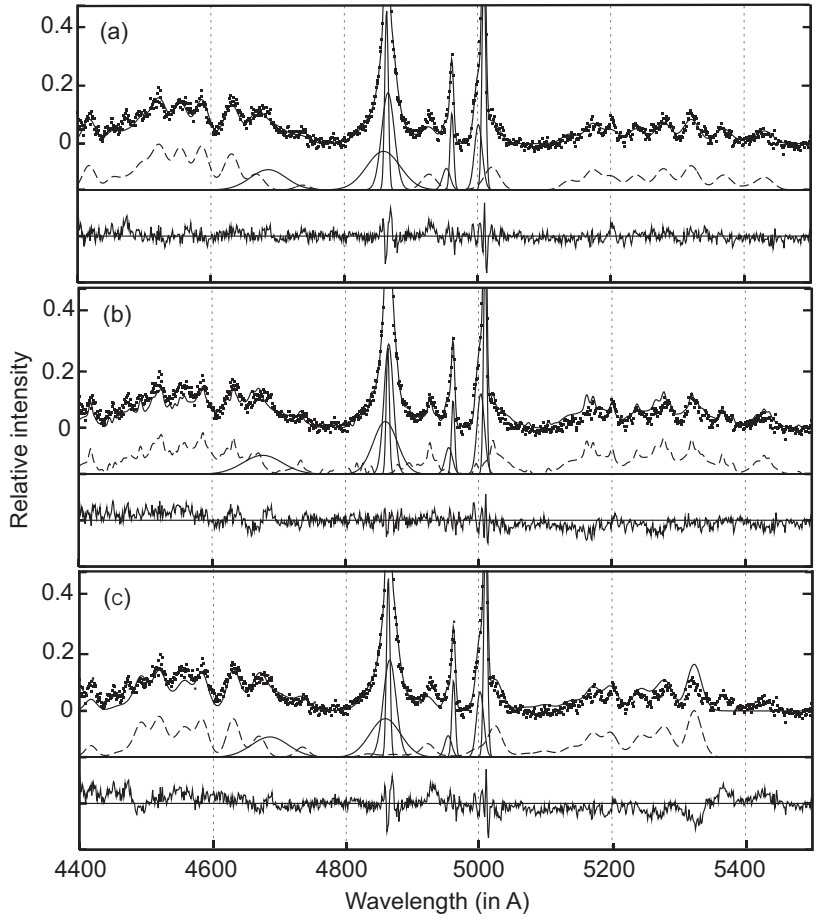}
\caption{Examples of fits to SDSS J$141755.54+431155.8$: with our template (a), with the empirical template of \citet{b302} (b), and with the theoretical template of \citet{b304} (c). Iron emission is a bit stronger in the blue bump, than in red. Our model fits this spectrum slightly better than other two.}
\label{6}
\end{figure*}

 We applied this template to our sample of 302 AGNs from SDSS database, and we compare it with other theoretical and empirical models (see Appendix B). We found that the template fit \ion{Fe}{2} emission very well (Figure \ref{5} and Figure \ref{6}). In the cases when the \ion{Fe}{2} emission of an object has different properties from I Zw 1, small disagreements are noticed in the lines for which the relative intensity was obtained from I Zw 1. However, the \ion{Fe}{2} emission in these objects shows larger disagreement with the empirical and theoretical models considered in Appendix B (see Figure \ref{7}). Although our \ion{Fe}{2} template has more free parameters than other templates used for fitting the \ion{Fe}{2} emission, it enables more precise fit of \ion{Fe}{2} lines in some AGNs (especially with very broad H$\beta$ line) than the other \ion{Fe}{2} templates we considered.

\subsection{Emission line decomposition and fitting procedure}

We assume that broad emission lines arise in two or more emission regions  \citep{a1,b31,b55,b00,b53}, so that their profiles are sums of Gaussians with different shifts, widths and intensities, which reflect the physical conditions of the emitting regions where the components arise.

For this reason, we fit the emission lines we considered ([\ion{O}{3}] $\lambda\lambda$4959, 5007 \AA, H$\beta$, He II $\lambda$4686 \AA, H$\alpha$ and [\ion{N}{2}] $\lambda\lambda$6548, 6583 \AA) with a sum of Gaussians, describing one Gaussian with 3 parameters (width, intensity and shift from transition wavelength).

Both lines of the [\ion{O}{3}] $\lambda\lambda$4959, 5007 \AA \ doublet originate from the same lower energy level and both have a negligible optical depth since the transitions are strongly forbidden. Taking this into account, we assumed that the [\ion{O}{3}] $\lambda$4959 \AA \ and [\ion{O}{3}] $\lambda$5007 \AA \ lines have the same emission-line profile. We fit each line of the doublet with either a single Gaussian, or, in the case of a significant asymmetry, with two Gausssians. The Gaussian that describes the [\ion{O}{3}] $\lambda$4959 \AA, has the same width and shift, as the one which describes [\ion{O}{3}] $\lambda$5007 \AA \ line, and we took their intensity ratio to be 2.99 \citep{b30}. [\ion{N}{2}] $\lambda\lambda$6548, 6583 \AA \ were fitted using the same procedure, with an intensity ratio of the doublet components of $\approx3$. The He II $\lambda$4686 \AA \ line was fitted with one broad Gaussian.

To fit the Balmer lines a number of Gaussian functions were used for each line. It is usually assumed that there are two components for H$\alpha$ and H$\beta$: a narrow component representing the NLR, and a broad one representing the BLR. However, the broad emission lines (BELs), are very complex and cannot be properly explained by single Gaussian (which would indicate an isotropic spherical region). Moreover, there are some papers  \citep{a1,b54,b31,b53,b55,b00,b52} where the broad lines are assumed to be emitted from two kinematicaly different regions: a ``Very Broad Line Region'' (VBLR), and an ``Intermediate Line Region'' (ILR).  We therefore tested fitting the broad lines with one and two Gaussians. We found that in most AGNs in our sample, the fit with two Gaussians was significantly better than the fit with one Gaussian.

\begin{figure*}

\includegraphics[width=0.37\textwidth]{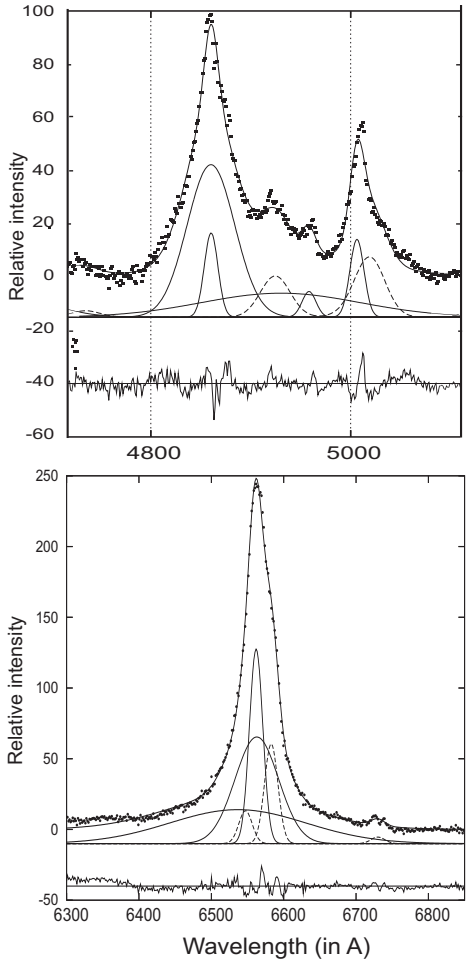}
\caption{Balmer lines in SDSS J$111603.13+020852.2$ spectrum. Top: H$\beta$ with a highly redshifted H$\beta$ VBLR component; bottom:  H$\alpha$ where the VBLR component is not redshifted.}
\label{8}
\end{figure*}

To apply a uniform model of the line fitting procedure, we assumed that Balmer lines have three components from the NLR, ILR and VBLR.  We exclude 4 spectra from the sample in which H$\alpha$ line could not be decomposed in this way, so finally our sample contains 133 spectra (from 137) with all Balmer lines which could be analyzed within the $\lambda\lambda$ 6400-6800 \AA \ range. To check this assumption on the rest of the spectra, we examined the corresponding correlations and we found that the luminosities and widths of the NLR, ILR and VBLR components of H$\alpha$ are highly correlated with the same parameters of the NLR, ILR and VBLR components of H$\beta$.  This is in favor of our three component decomposition (Table \ref{tbl-3}). For the 133 objects, which have both H$\beta$ and H$\alpha$ lines, it can be noticed that 16 spectra have large redshifts of the VBLR H$\beta$ component relative to VBLR of H$\alpha$, which causes disagreement and reduces the correlation between the VBLR shifts of the rest of objects. Without these 16 objects, H$\beta$ and H$\alpha$ VBLR components correlate in width and shift, with coefficient of correlation $\sim0.40$, P $<$ 0.0001. An example of shifts between the H$\alpha$ and H$\beta$ VBLR components is shown in Figure \ref{8}. The AGN SDSS J$111603.13+020852.2$ has its VBLR H$\beta$ component redshifted by $\approx$4258 $\mathrm{kms^{-1}}$, while shift of the H$\alpha$ VBLR component is $\approx$-1200 $\mathrm{kms^{-1}}$.

We assume that all narrow lines (and narrow components of the broad lines), originate from the same NLR, and thus expect that these lines will have the same shifts and widths. We therefore took the Gaussian parameters of the shift and width of [\ion{O}{3}], [\ion{N}{2}], and the NLR components of H$\beta$ and H$\alpha$ to have the same values.

\begin{figure*}

\includegraphics[width=0.50\textwidth,angle=270]{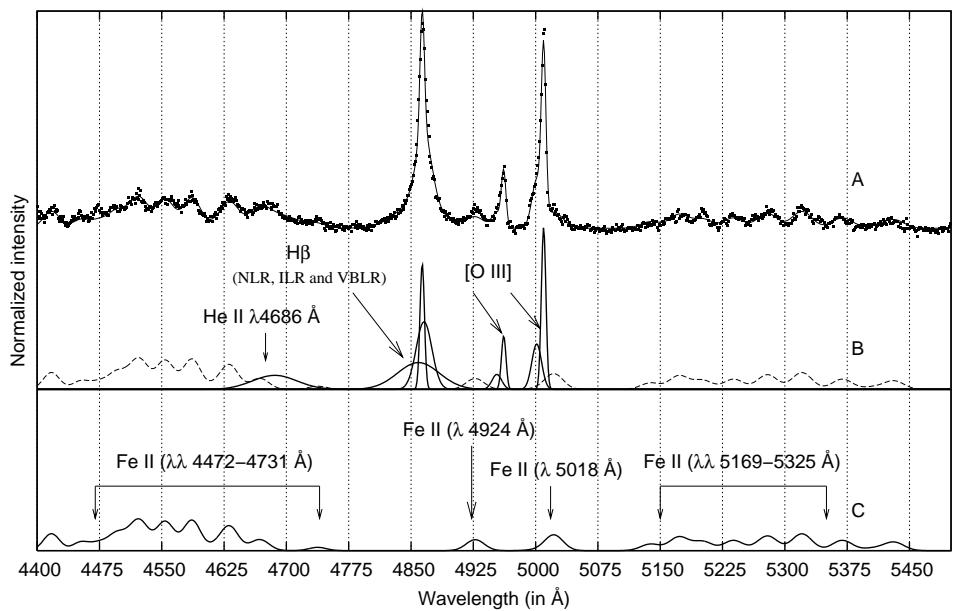}
\caption{The spectrum of SDSS J$141755.54+431155.8$ in the $\lambda\lambda$ 4400-5500 \AA \ region: (A) the observed spectra (dots) and the best fit (solid line). (B) H$\beta$ fit with the sum of three Gaussians representing emission from the NLR, ILR and VBLR. The [\ion{O}{3}] $\lambda\lambda$4959, 5007 \AA \ lines are fit with two Gaussians for each line of the doublet and He II $\lambda$4686 \AA \ is fit with one broad Gaussian. The \ion{Fe}{2} template is denoted with a dashed line, and also represented separately in panel C (bottom).}
\label{9}
\end{figure*}

\begin{figure*}

\includegraphics[width=0.50\textwidth]{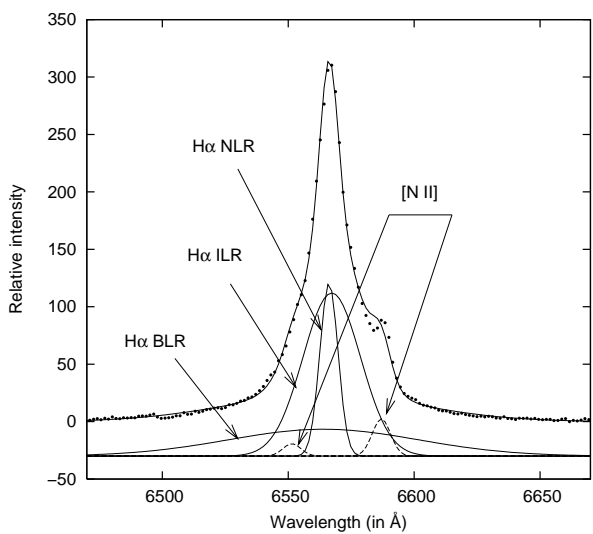}
\caption{Fits of SDSS J$141755.54+431155.8$ in the $\lambda\lambda$ 6400-6800 \AA \ region: H$\alpha$ is fit with the sum of three Gaussians which represent emission from  NLR, ILR and VBLR and [\ion{N}{2}] $\lambda\lambda$6548, 6583 \AA \ lines are fit with one Gaussian for each line of doublet. Narrow [\ion{N}{2}] lines are denoted with dashed lines.}
\label{10}
\end{figure*}

We separately fit lines from the $\lambda\lambda$ 4400-5500 \AA \ range (\ion{Fe}{2} lines, [\ion{O}{3}], He II and H$\beta$) with 23 free parameters, and lines from the $\lambda\lambda$ 6400-6800 \AA \ range (H$\alpha$ and [\ion{N}{2}]) with 8 free parameters. We applied a ${\chi}^2$ minimization routine \citep{b31}, to obtain the best fit. Examples of an AGN spectrum fit in both ranges are shown in Figure \ref{9} and Figure \ref{10}.

\subsection{Line parameters}

\begin{figure*}

\includegraphics[width=0.5\textwidth]{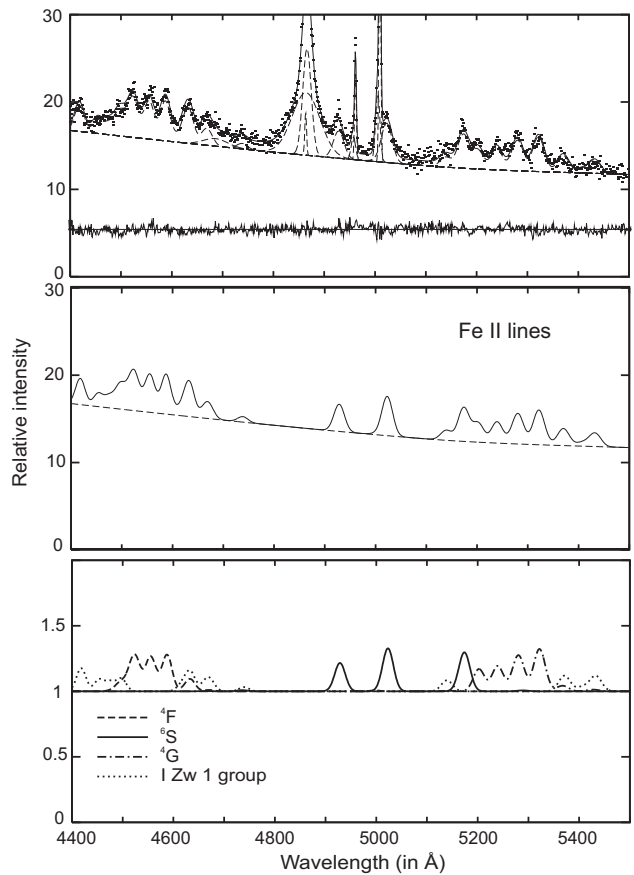}
\caption{Example of measuring of equivalent widths of the \ion{Fe}{2} group of lines.  Top: fit spectrum (SDSS J020039.15-084554.9); middle: \ion{Fe}{2} lines and continuum; bottom: \ion{Fe}{2} lines normalized on the continuum level. }
\label{11}
\end{figure*}

\begin{figure*}

\includegraphics[width=0.45\textwidth]{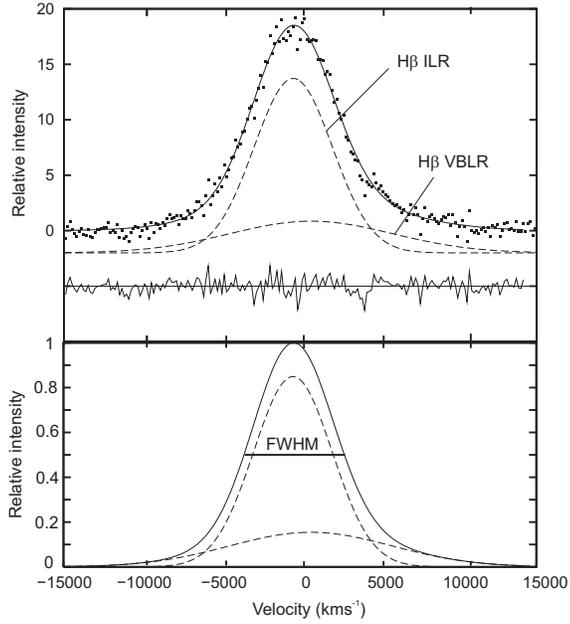}
\caption{Example of measuring of FWHM; Top: fitted broad H$\beta$ (ILR+VBLR); bottom: FWHM is obtained as width on half maximum of the sum of ILR and VBLR Gaussians. }
\label{11_2}
\end{figure*}

We compared the shifts, widths, equivalent widths, and luminosities of the lines. The shifts and widths were obtained directly from the fit (parameters of width and shift).  Luminosities were calculated using the formulae given in \citet{b306}, with adopted cosmological parameters:  $\Omega_M$ = 0.27, $\Omega_\Lambda$ = 0.73 and $\Omega_k$ = 0. We adopt a Hubble constant, $\mathrm{H_o=71\ \rm kms^{-1}Mpc^{-1}}$.
 The continuum luminosity was obtained from the average value of continuum flux measured between 5100 \AA \ ($\lambda\lambda$5095-5100 \AA \ and $\lambda\lambda$5100-5105 \AA).   Equivalent widths were measured with respect to the continuum below the lines. The continuum was estimated by subtracting all fitted lines from the observed spectra. Then, the line of interest (or the \ion{Fe}{2} template), which is fit separately, is added to the continuum. Equivalent widths are obtained by normalizing the lines on continuum  level, and measuring their fluxes (see Figure \ref{11}).  To determine the full width at half maximum (FWHM) of the broad component of the Balmer lines, the VBLR and ILR components from the fit are considered as it is shown in Figure \ref{11_2}. The broad Balmer line (VBLR+ILR) is then normalized on unity, and the FWHM is measured at half of the maximum intensity.

\section{Results}

\subsection{The ratios of \ion{Fe}{2} line groups vs. other spectral properties}

Some objects have relative intensities of \ion{Fe}{2} lines similar to I Zw 1, while others show significant disagreement, which usually shows up as  stronger iron emission in the blue bump (mainly $F$ lines) than one in the red bump (mainly $G$ lines), comparing to  I Zw 1. That objects could be fitted well with the template model which assumes that relative intensities between the \ion{Fe}{2} lines from different line groups ($F$, $S$ and $G$) are  different for different objects.

\begin{figure*}

\includegraphics[width=0.32\textwidth]{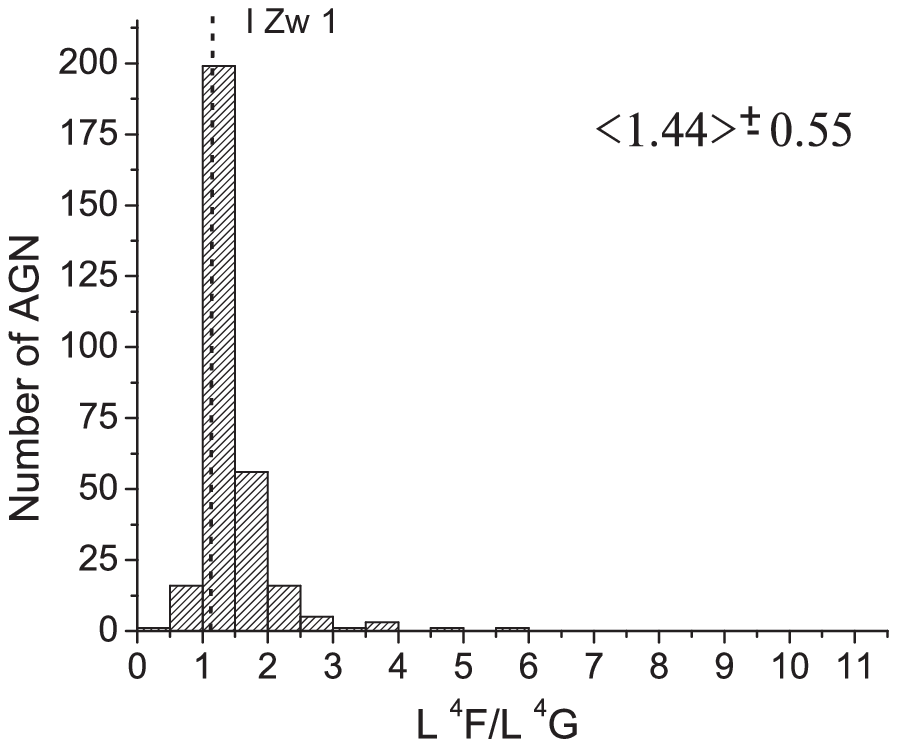}
\includegraphics[width=0.32\textwidth]{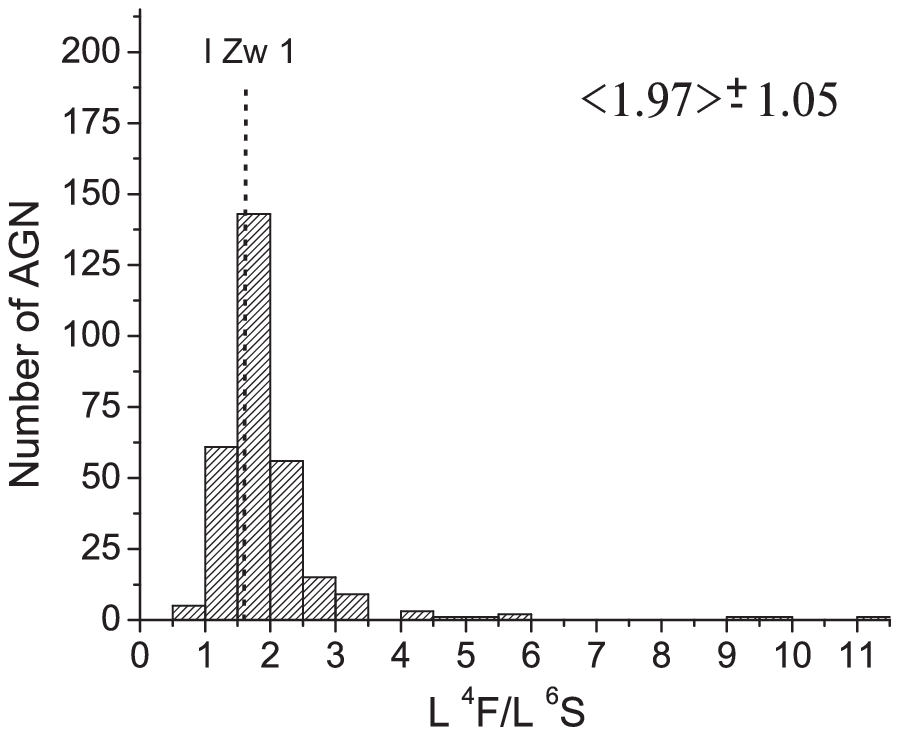}
\includegraphics[width=0.32\textwidth]{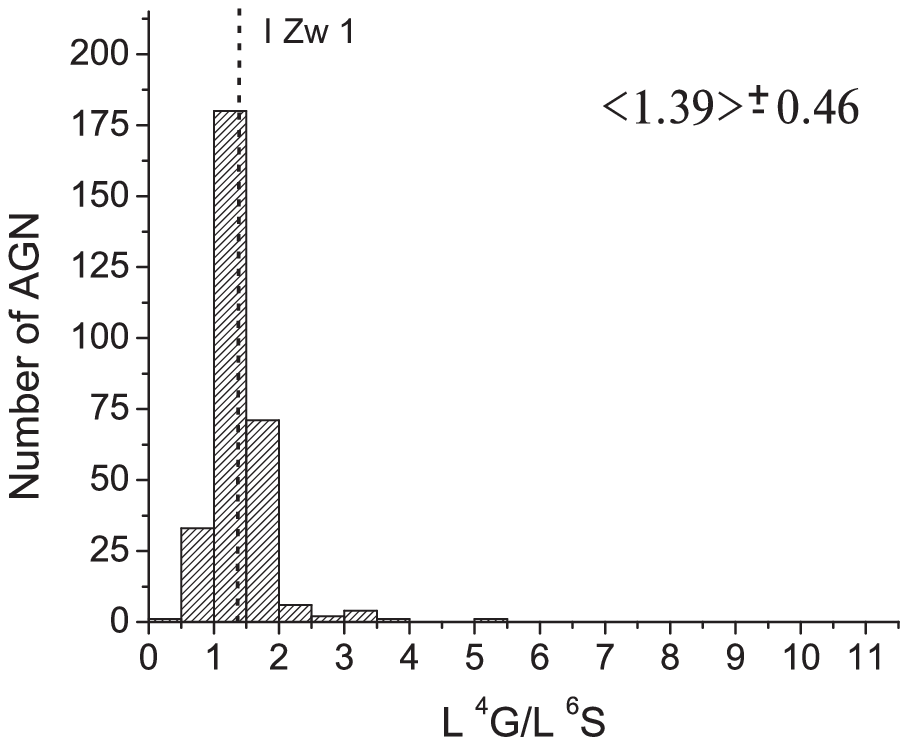}
\caption{ Distribution of the ratios of luminosities of the three \ion{Fe}{2} groups ($F$, $S$, and $G$).  The mean value and the dispersion of the distribution are given in the plots. We indicate with a dashed vertical line the ratio of \ion{Fe}{2} groups we obtained in the I Zw 1 spectrum. }
\label{12}
\end{figure*}

Difference between the relative intensities of the \ion{Fe}{2} lines from different parts of iron shelf (blue, red and central) can be illustrated well with histograms of distribution of the \ion{Fe}{2} group ratio ($F$/$G$, $F$/$S$ and $G$/$S$) for the sample (Figure \ref{12}). The average value of the $F/G$ luminosity ratio is 1.44$\pm$0.55, of the $F/S$ ratio 1.97$\pm$1.05, and of the $G/S$ ratio 1.39$\pm$0.46\footnote{ Here we give the averaged ratio value and dispersion.}.  We found that the majority of objects have \ion{Fe}{2} group ratios close to the average values, but still significant number of objects are showing difference since the intensity ratios of \ion{Fe}{2} line groups vary in different objects (see Figure \ref{12}). The ratios of the \ion{Fe}{2} line groups in I Zw 1 are indicated by the vertical dashed lines ($F/G = 1.10$, $F/S = 1.56$ and $G/S = 1.41$).

\begin{figure*}

\includegraphics[width=0.32\textwidth]{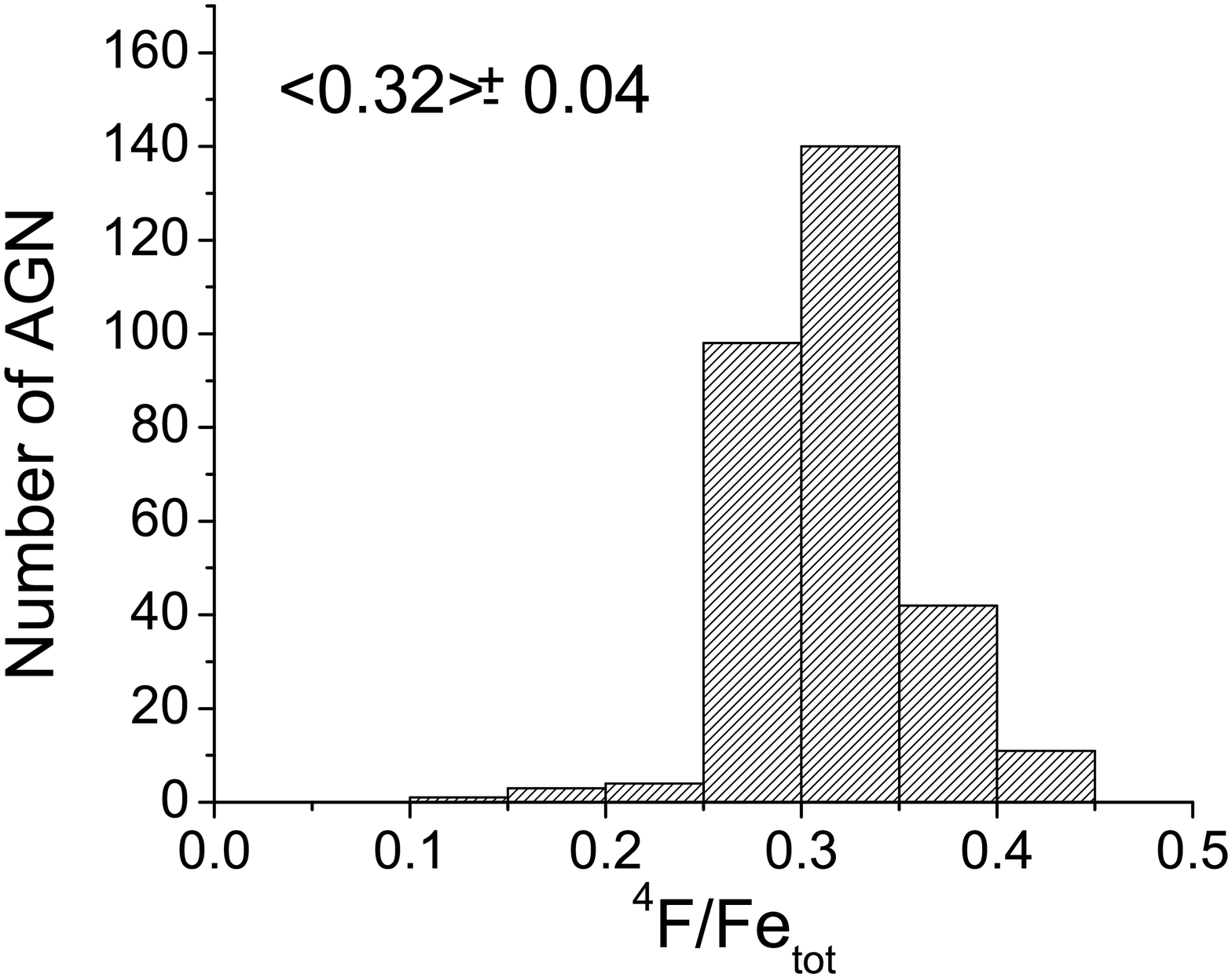}
\includegraphics[width=0.32\textwidth]{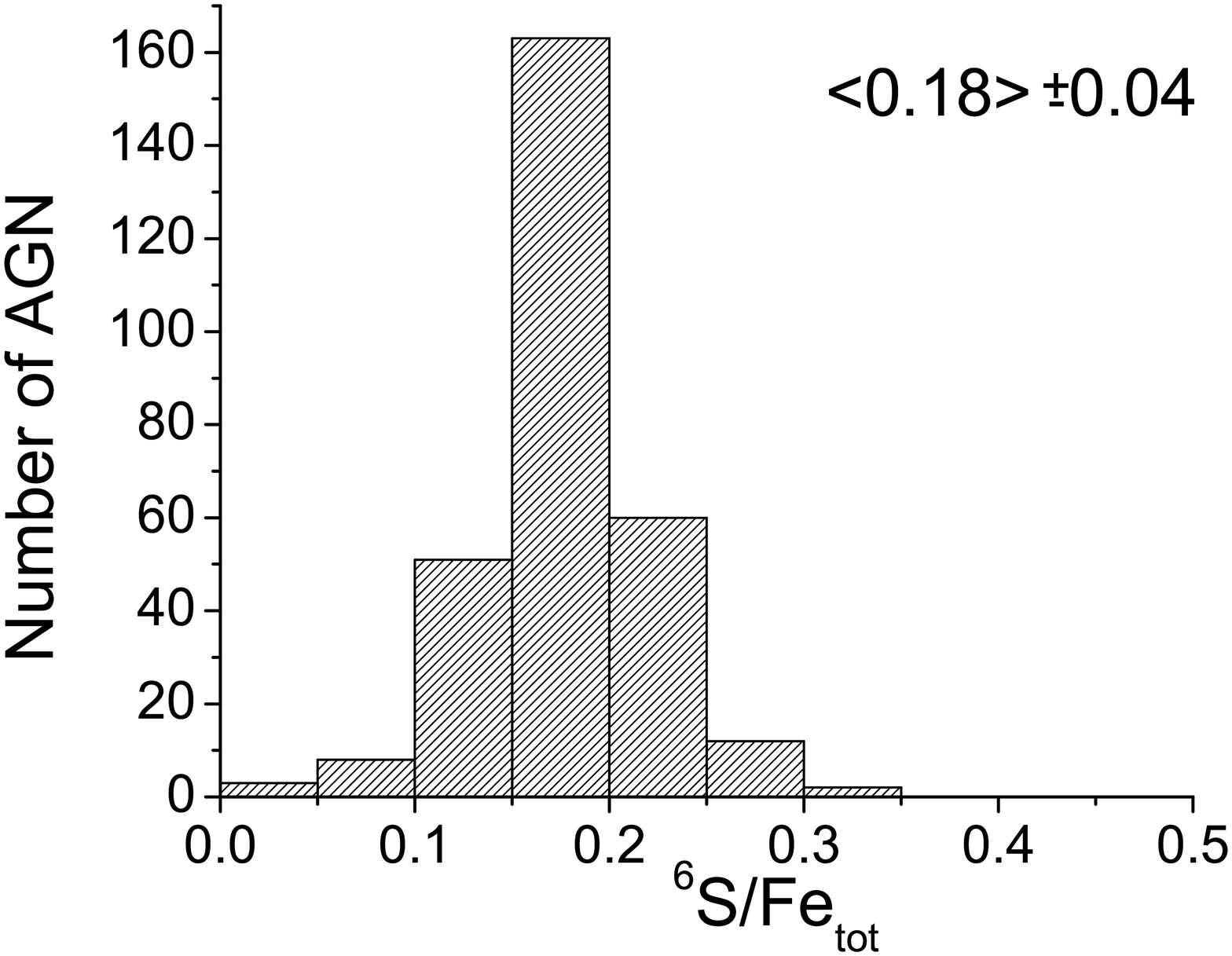}
\includegraphics[width=0.32\textwidth]{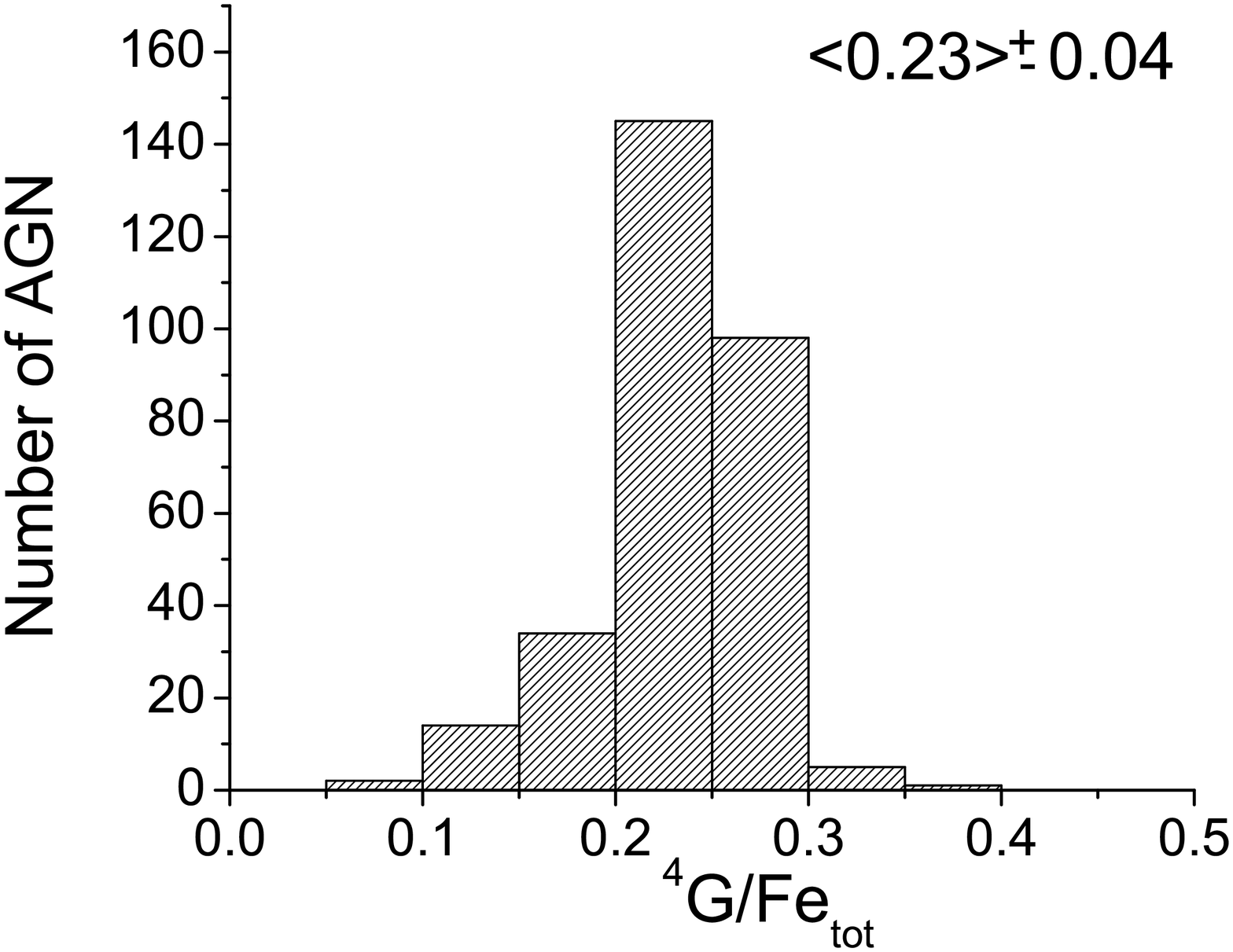}
\caption{The distribution of the ratios of total \ion{Fe}{2} (4400-5500 \AA) and \ion{Fe}{2} groups ($F$, $S$ and $G$). The mean value and the dispersion of the distribution are shown.}
\label{fetot}
\end{figure*}

We analyzed the flux ratios of the \ion{Fe}{2} multiplet groups and the total \ion{Fe}{2} flux (see histograms in Figure \ref{fetot}), and we found that lines from multiplets 37 and 38 (group $F$) generally have the largest contribution to the total \ion{Fe}{2} emission within the $\lambda\lambda$4400-5500 \AA \ range (about 32\%), while group $S$ contributes about 18\% and $G$ about 23\%. As expected, there are very strong correlations between the equivalent widths of \ion{Fe}{2} line groups (Table \ref{tbl-4}).

\begin{figure*}

\includegraphics[width=0.32\textwidth]{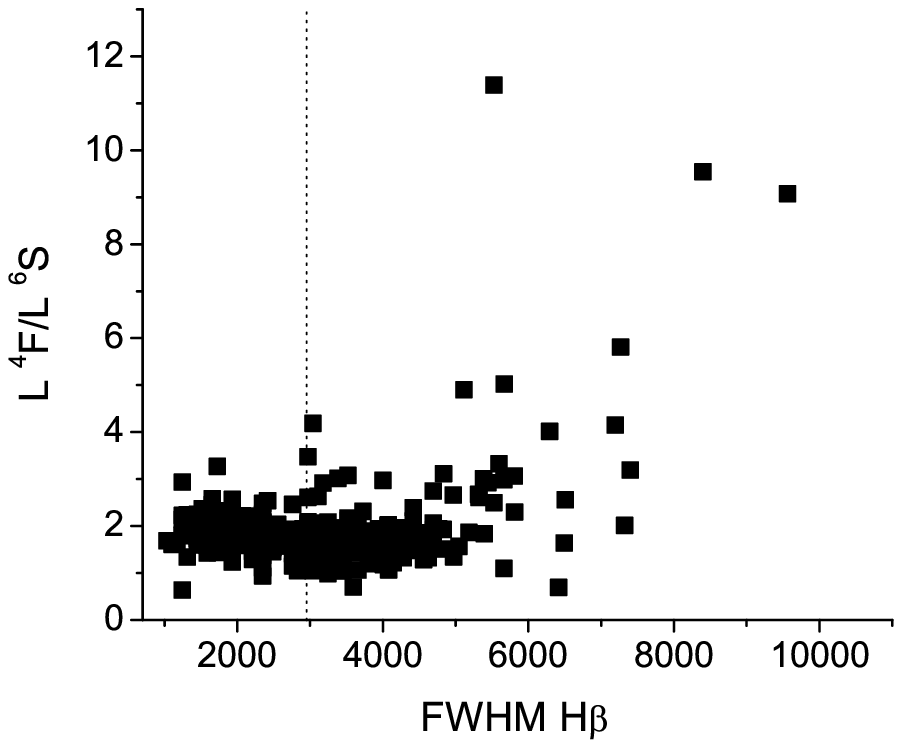}
\includegraphics[width=0.32\textwidth]{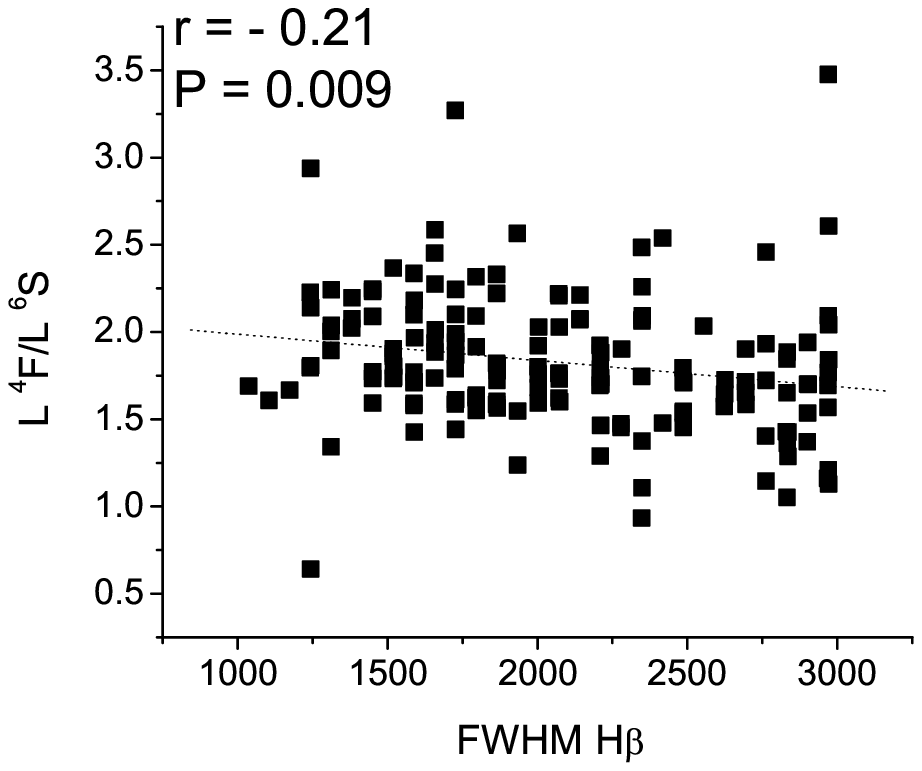}
\includegraphics[width=0.32\textwidth]{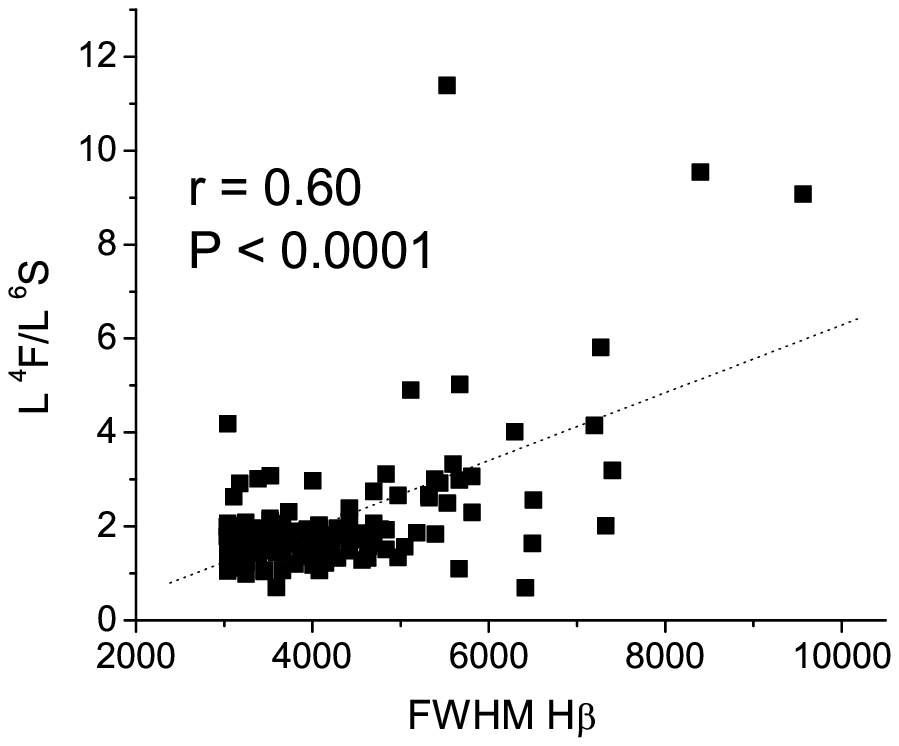}
\caption{The relationship between $F/S$ and H$\beta$ FWHM, for total sample (left), sub-sample of H$\beta$ FWHM $<$ 3000 $\mathrm{kms^{-1}}$ (middle) and the sub-sample of H$\beta$ FWHM $>$ 3000 $\mathrm{kms^{-1}}$ (right).}
\label{13}
\end{figure*}

Since line intensities and their ratios are indicators of physical properties of the plasma where those lines arise, we have investigated relations among the ratios of \ion{Fe}{2} line groups with various spectral properties.  A correlation between the FWHM H$\beta$ and the EW \ion{Fe}{2} emission has been found previously \citep{b20,b9,b0001}. Consequently, one can expect that FWHM H$\beta$ may be connected with the \ion{Fe}{2} line group ratios. We therefore investigated correlations between the ratios of the \ion{Fe}{2} groups ($F/G$, $F/S$, and $G/S$) and H$\beta$ FWHM. We excluded the three outliers, which have negligible flux for $S$ group. As it can be seen in Figure \ref{13} (first panel) there are different trends for spectra with FWHM H$\beta$ less than and greater than $\sim$3000 $\mathrm{kms^{-1}}$. Therefore, we divided the sample into two sub-samples: 158 objects with FWHM H$\beta$ $<$3000 $\mathrm{kms^{-1}}$ and 141 objects with FWHM H$\beta$ $>$3000 $\mathrm{kms^{-1}}$ (see Figure \ref{13} and Table \ref{tbl-5}). A similar subdivision of AGN samples was performed by \citet{b000,b0001}, but for FWHM H$\beta\approx 4000 \ \rm kms^{-1}$ as the limiting FWHM of the subdivision. We found that FWHM H$\beta\approx 3000 \ \rm kms^{-1}$ is more a appropriate place to divide the objects into two groups, since all objects with strongly redshifted VBLR H$\beta$ component (see Figure \ref{8}) belong to the FWHM H$\beta$ $>$3000 $\mathrm{kms^{-1}}$ subsample. \citet{b000} also noticed that AGN with FWHM H$\beta$ $>$4000 $\mathrm{kms^{-1}}$, have a larger redshifted VBLR H$\beta$ ($\sim$5000 $\mathrm{kms^{-1}}$).

 In Table \ref{tbl-5} we present the correlations between the luminosity ratios of the \ion{Fe}{2} line groups and other spectral properties such as: FWHM H$\beta$, Doppler width of H$\beta$ broad components, width and shift of iron lines and continuum luminosity ($\mathrm{L_{\lambda5100}}$). They are examined for the total sample and for two sub-samples, divided according to FWHM H$\beta$, and denoted in Table \ref{tbl-5} as (1) for FWHM H$\beta$ $<$3000 $\mathrm{kms^{-1}}$ and (2) for H$\beta$ FWHM $>$3000 $\mathrm{kms^{-1}}$.

We found a difference in the correlations for those two sub-samples. For the H$\beta$ FWHM $>$ 3000 $\mathrm{kms^{-1}}$ sub-sample (2) all three ratios ($F/G$, $F/S$ and $G/S$) increase as H$\beta$ width increases. On the other hand, for the H$\beta$ FWHM $<$ 3000 $\mathrm{kms^{-1}}$ sub-sample no correlation is observed between these properties.

The observed correlation between the $G/S$ group ratio and H$\beta$ FWHM may be caused by intrinsic reddening since reddening increases as objects get more edge-on \citep{gas} and the H$\beta$ FWHM also increases \citep{b26}. However, this cannot explain the $F/G$ and $F/S$ correlation with H$\beta$ FWHM, and also, a strong influence of reddening on the $G/S$ ratio is not expected because of the close wavelengths of lines from those groups.

We also studied correlations between continuum luminosity ($\mathrm{L_{\lambda5100}}$) and \ion{Fe}{2} group ratios, since these may indicate excitation processes in the emitting region. We found that the $F/G$ and $F/S$ ratios decrease in objects where the continuum level is higher. Observed correlations are stronger for H$\beta$ FWHM $<$ 3000 $\mathrm{kms^{-1}}$ sub-sample ($F/G$ vs. $\mathrm{L_{\lambda5100}}$: $r = - 0.51$, $P < 0.0001$ and $F/S$ vs. $\mathrm{L_{\lambda5100}}$: $r = - 0.41$, $P < 0.0001$). The ratio of $G/S$ seems not to be dependent on continuum luminosity.

 These correlations are the opposite of what would be expected from reddening, since reddening decreases with luminosity \citep{gas}. There are a few effects that could destroy expected reddening effect: (i) incorrect continuum subtraction, (ii) host galaxy fraction which depends on the Eddington ratio and (iii) different contribution of starlight in SDSS fiber which depends on the luminosity of the host galaxy and the redshift \citep[see][]{GK}. There is a small possibility that these effects are important. As we already noted, the error-bar in the continuum subtraction is smaller than 5\%. Also, we examined if the host galaxy contribution have influence on considered correlations. We found no correlations between the host galaxy fraction (determined in $\lambda$5100 \AA) and \ion{Fe}{2} group ratios.\footnote{The amount of stellar continuum is taken from paper \citet{VB} for 106 common objects (see Appendix A).} The correlations appear instead to be connected with a weaker Baldwin effect for F group (see Table \ref{tbl-7}).

\begin{figure*}

\includegraphics[width=0.32\textwidth]{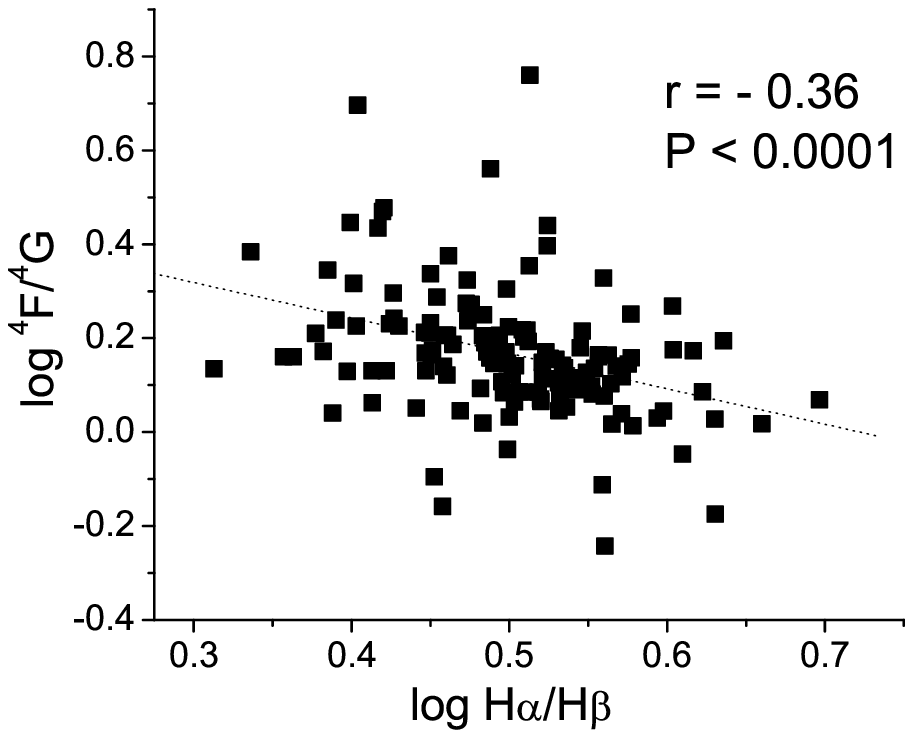}
\includegraphics[width=0.32\textwidth]{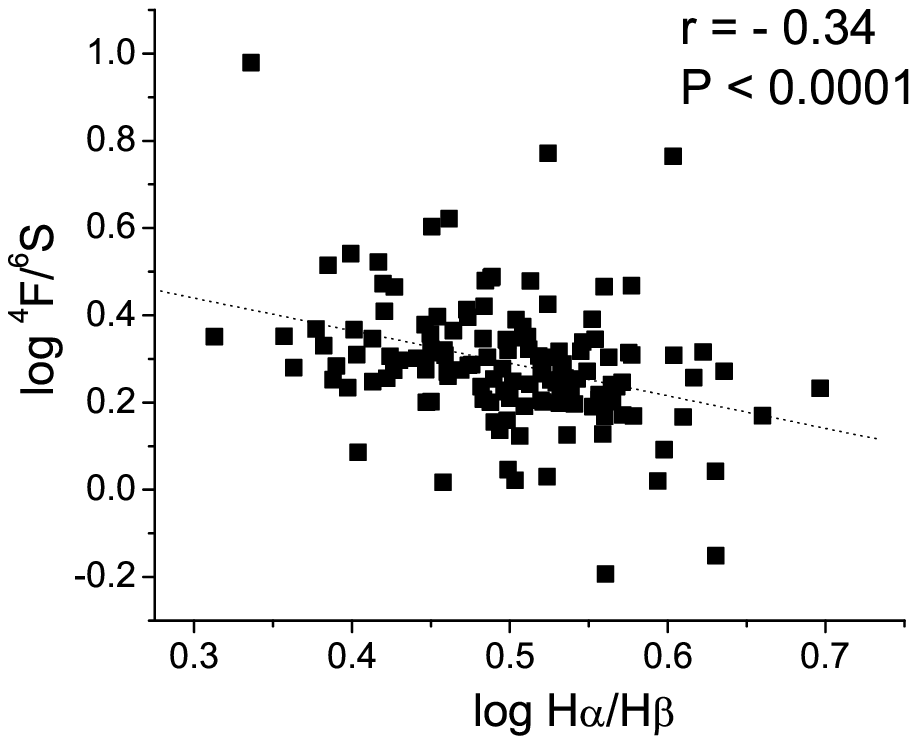}
\includegraphics[width=0.32\textwidth]{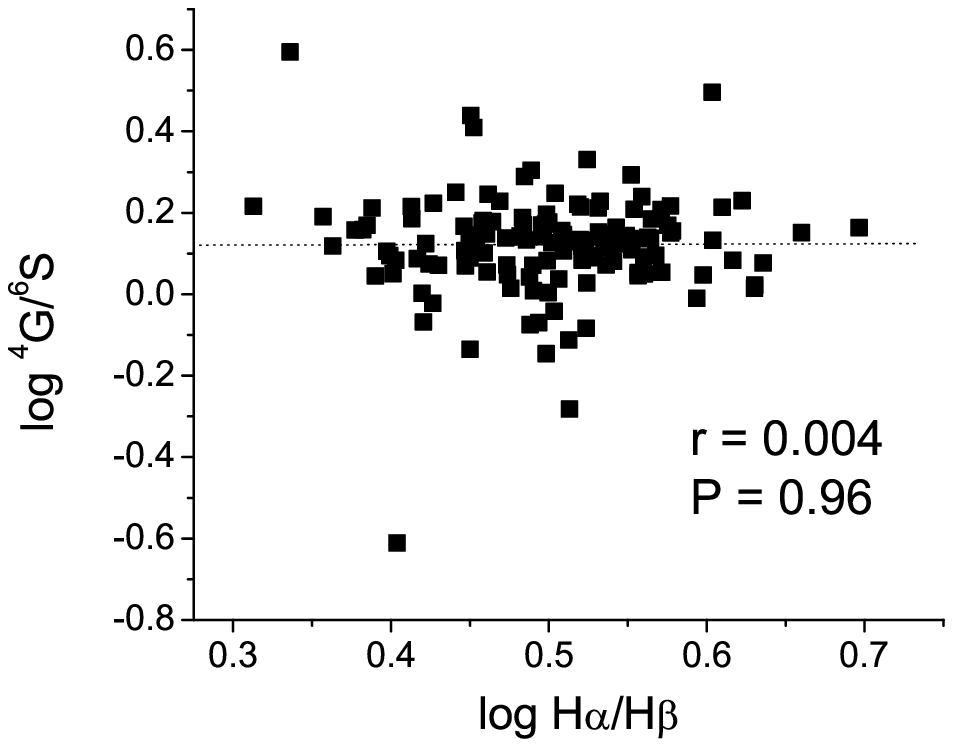}
\caption{The relationship between \ion{Fe}{2} group ratios and the Balmer decrement.}
\label{logab}
\end{figure*}

We also investigated the correlation of the log(L H$\alpha$/L H$\beta$) with ratios of \ion{Fe}{2} groups. We found correlations with $F/G$ ($r= - 0.36$, $P < 0.0001$) and with $F/S$ ratio ($r= - 0.34$, $P < 0.0001$), but there is no correlation with $G/S$ (Figure \ref{logab}, Table \ref{tbl-6}), i.e., between ratio of red and central part.  There is possibility that these correlations are caused by intrinsic reddening. 

\subsection{Connection between kinematical properties of \ion{Fe}{2} lines and Balmer lines (H$\alpha$ and H$\beta$)}

We assume that broadening of the lines arise by Doppler effect caused by random velocities of emission clouds, and that shifts of lines are a consequence of systemic motions of the emitting gas. We therefore studied the kinematical connection among emission regions by analyzing relationships  between their widths and shifts obtained from our best fits.

\begin{figure*}

\includegraphics[width=0.45\textwidth]{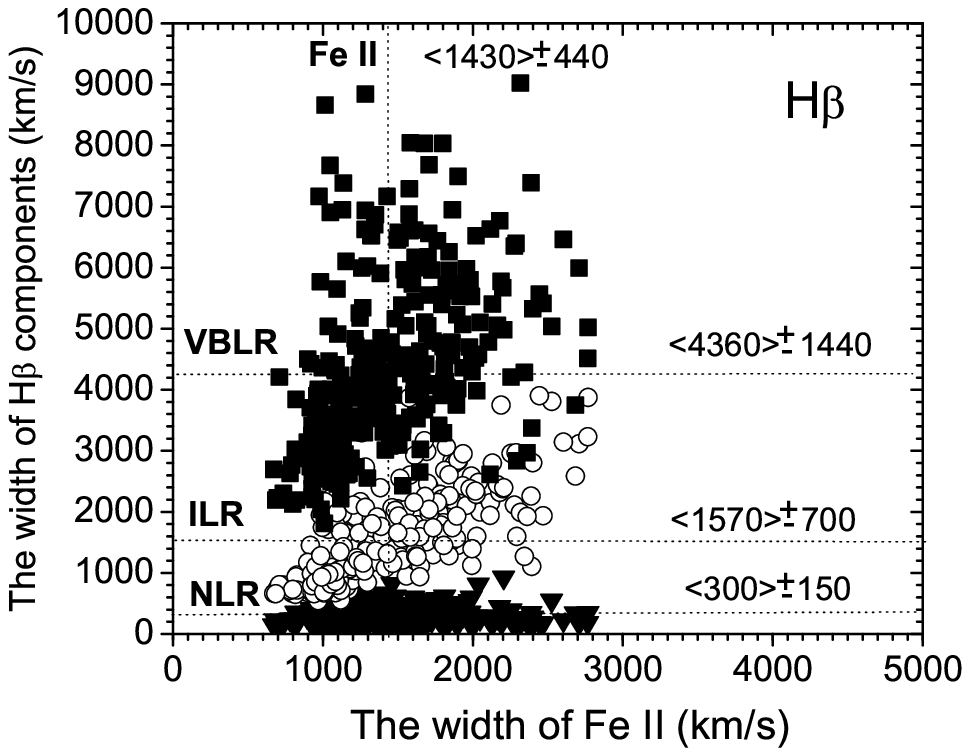}
\includegraphics[width=0.45\textwidth]{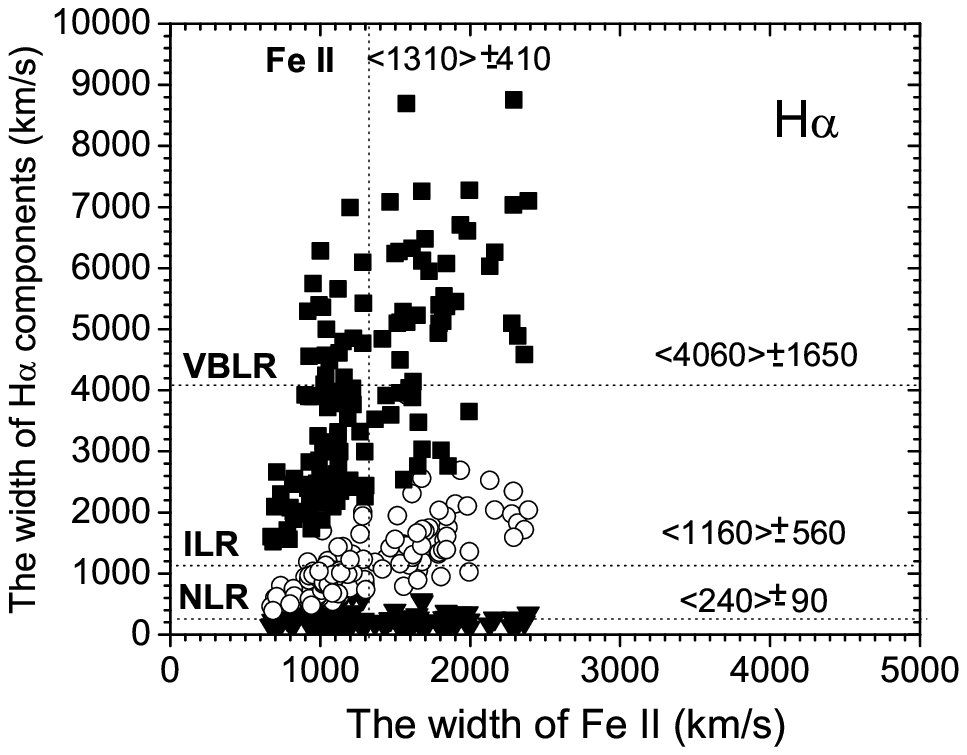}
\caption{ The widths of the \ion{Fe}{2} lines compared to the widths of the H$\beta$ (left) and H$\alpha$ (right) components. On the X-axis are the widths of \ion{Fe}{2}, and on Y-axis are the widths of the NLR (triangles), ILR (circles), and VBLR (squares) components of H$\beta$ (H$\alpha$). The dotted vertical line shows the average value of \ion{Fe}{2} widths, while the dotted horizontal lines show the average values of the H$\beta$ (H$\alpha$) components. The average value of for the \ion{Fe}{2} lines is the same as or very close to the average width value of the ILR components of H$\beta$ and H$\alpha$.}
\label{14}
\end{figure*}

The Doppler widths of \ion{Fe}{2} and Balmer lines (H$\beta$ and H$\alpha$) are compared in Figure \ref{14}. On X-axis we present \ion{Fe}{2} width and on the Y-axis the widths of the H$\beta$ (first panel) and H$\alpha$ (second panel) components. The widths of the NLR components are denoted by triangles, the ILR components with circles, and the VBLR components with squares. Dotted lines show the average values of the widths: vertical lines for \ion{Fe}{2} components and horizontal lines for H$\beta$ (or H$\alpha$). For the sample of 302 AGN (first panel) the average value of \ion{Fe}{2} width  and dispersion of the sample are $1430\pm440$ $\mathrm{kms^{-1}}$, while the average values for H$\beta$ components are: $300\pm150$ $\mathrm{kms^{-1}}$ (NLR), $1570\pm700$ $\mathrm{kms^{-1}}$ (ILR) and $4360\pm1440$ $\mathrm{kms^{-1}}$ (VBLR). For the selected sample of 133 AGNs which have the H$\alpha$ line in the spectra (second panel), the averaged value of the \ion{Fe}{2} width is $1310\pm410$ $\mathrm{kms^{-1}}$ and for the H$\alpha$ components: $240\pm90$ $\mathrm{kms^{-1}}$ (NLR), $1160\pm560$ $\mathrm{kms^{-1}}$ (ILR) and $4060\pm1650$ $\mathrm{kms^{-1}}$ (VBLR). It is obvious that the averaged \ion{Fe}{2} width ($\sim1400$ $\mathrm{kms^{-1}}$) is very close to the average widths of the ILR H$\beta$ and H$\alpha$ components (1568 $\mathrm{kms^{-1}}$ and 1156 $\mathrm{kms^{-1}}$ respectively), while the averaged widths of NLR and VBLR components are significantly different from the \ion{Fe}{2} width.

\begin{figure*}

\includegraphics[width=0.32\textwidth]{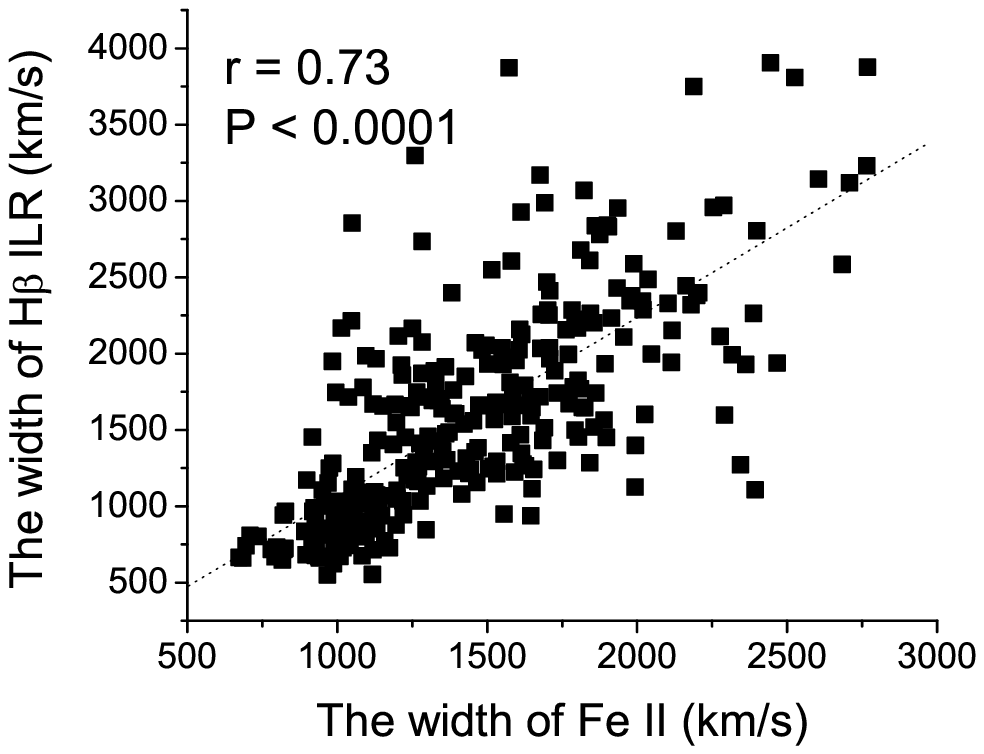}
\includegraphics[width=0.32\textwidth]{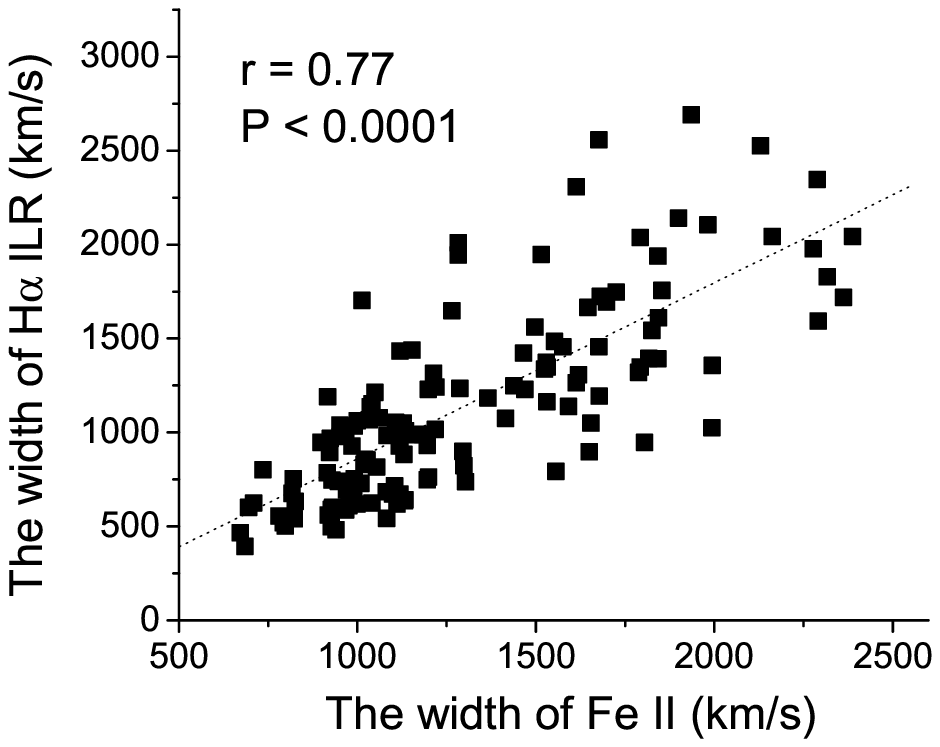}
\includegraphics[width=0.32\textwidth]{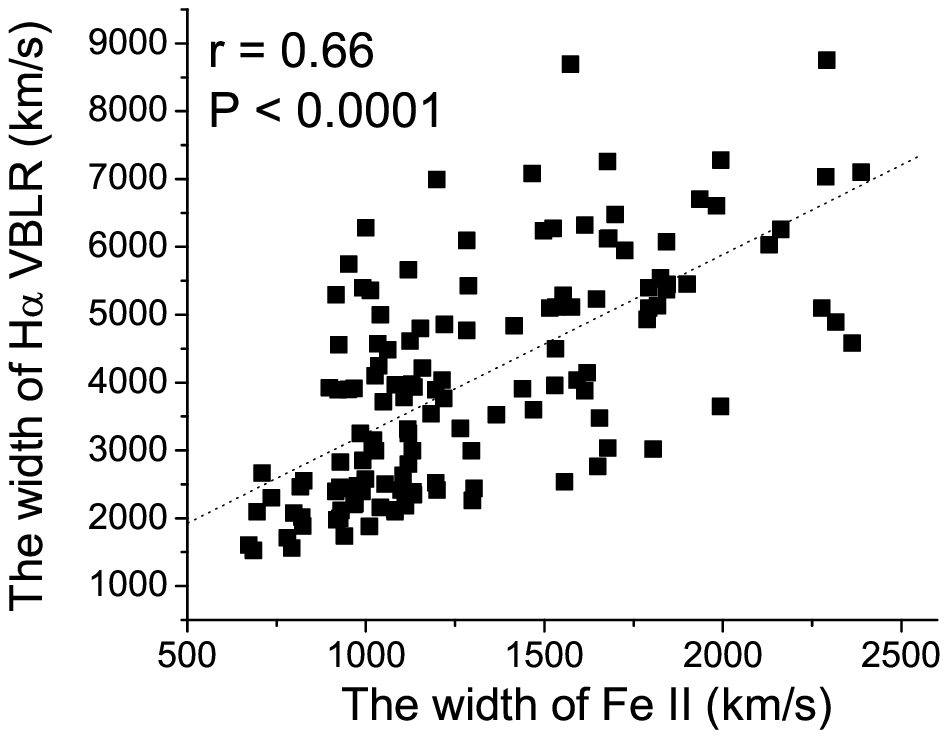}
\caption{ The correlation between the widths of \ion{Fe}{2} and the H$\beta$ ILR (left), H$\alpha$ ILR (middle) and H$\alpha$ VBLR components (right). In all cases, correlations are observed ($r = 0.67$, 0.72, 0.62), indicating kinematical connection between these emission regions.}
\label{15}
\end{figure*}

Relationships among the widths of the \ion{Fe}{2} and ILR components are also presented in Figure \ref{15} (Table \ref{tbl-7}). The correlation between the \ion{Fe}{2} width and the width of H$\alpha$ ILR is $r=0.77$, $P < 0.0001$, and between the \ion{Fe}{2} width and the H$\beta$ ILR width it is $r=0.73$, $P < 0.0001$. We also found a correlation between the widths of \ion{Fe}{2} and the VBLR ($r=0.66$, $P < 0.0001$ for the H$\alpha$ VBLR, and r=0.45, $P<0.0001$ for H$\beta$ VBLR).

Relationships between the shifts of \ion{Fe}{2} and H$\alpha$ and H$\beta$ components were also considered (Table \ref{tbl-7}). We found correlations with the shifts of the H$\alpha$ ILR ($r=0.30$, $P=0.0004$) and the H$\beta$ ILR component ($r=0.39$, $P<0.0001$), but no correlation between the shift of \ion{Fe}{2} and the shifts of other H$\alpha$ and H$\beta$ components.

We found that \ion{Fe}{2} lines tend to have an averaged redshift of 270$\pm$180 $\mathrm{kms^{-1}}$  with respect to the transition wavelength, and with respect to narrow lines 100$\pm$240 $\mathrm{kms^{-1}}$ (see Figure \ref{18}).

\subsection{Correlations between the \ion{Fe}{2} line groups and other emission lines}

 Correlations between the luminosity of the \ion{Fe}{2} multiplet groups, total \ion{Fe}{2} and luminosities of the [\ion{N}{2}] and [\ion{O}{3}] lines are presented in Table \ref{tbl-8}. In the $L$ [\ion{N}{2}] vs. $L$ \ion{Fe}{2} plot, seven outliers are observed with negligible [\ion{N}{2}].  They are probably caused by an underestimate in the fit of the [\ion{N}{2}] lines due to the blending with the H$\alpha$ line. Without these outliers, correlation is $\sim0.70$, P$<$0.0001. There is no correlation between $L$ [\ion{N}{2}]/$L$ [\ion{O}{3}] vs. $L$ \ion{Fe}{2}.

We analyzed relationships among the luminosities of the \ion{Fe}{2} line groups and the NLR, ILR and VBLR H$\alpha$ components (Table \ref{tbl-8}). All three H$\alpha$ components are correlated with the \ion{Fe}{2} line groups. Correlations are stronger with the ILR and VBLR components than with the NLR components. The same analysis was carried out for the luminosities of \ion{Fe}{2} line groups and the NLR, ILR and VBLR components of H$\beta$ (Table \ref{tbl-8}). NLR component of H$\beta$ also has a weaker correlation with \ion{Fe}{2} lines (r$\sim$0.60, P$<$0.0001) than broad H$\beta$ components (r$\sim$0.90, P$<$0.0001).

\begin{figure*}
\includegraphics[width=0.37\textwidth]{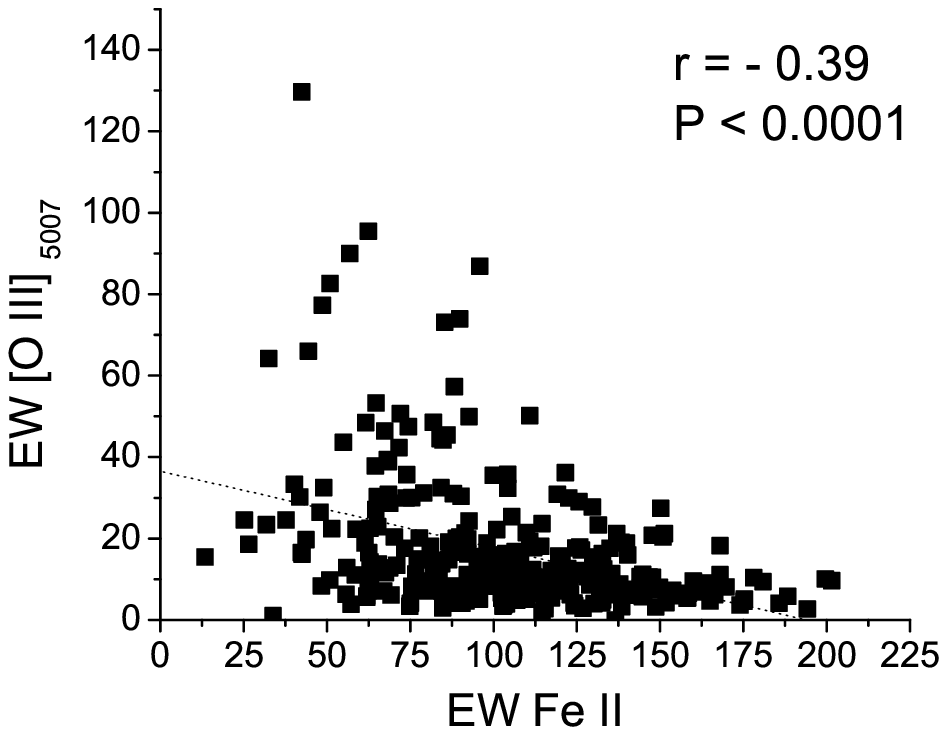}
\includegraphics[width=0.37\textwidth]{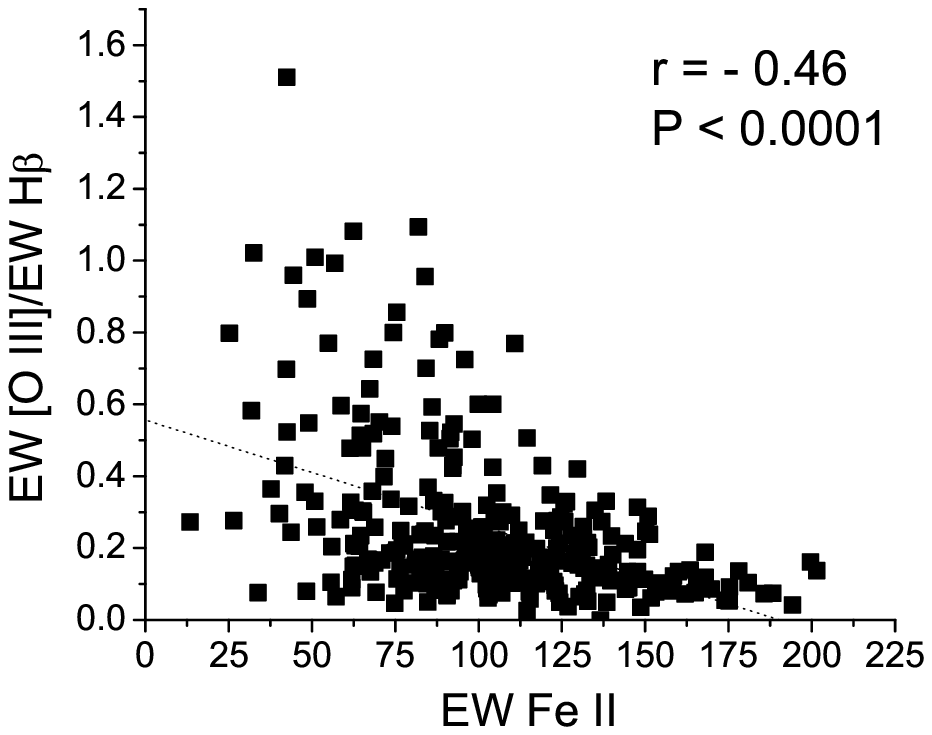}
\caption{Relationship between the EW [\ion{O}{3}] $\lambda$5007 \AA \ vs. EW \ion{Fe}{2} (left) and EW [\ion{O}{3}]/EW H$\beta$ vs. EW \ion{Fe}{2} (right).}
\label{17}
\end{figure*}

We also investigated the anti-correlations between EW \ion{Fe}{2} and EW [\ion{O}{3}], and between EW \ion{Fe}{2} and EW [\ion{O}{3}]/EW H$\beta$ \citep{b9}. We confirmed the existence of these relationships in our sample, with \ion{Fe}{2} lines separated into $F$, $S$, and $G$ line groups (Figure \ref{17}, Table \ref{tbl-9}).  A difference from \citet{b9}, who measured the equivalent widths of all lines referring to the continuum level at the $\lambda4861$ \AA, is that we calculated the equivalent widths by estimating the continuum below all the \ion{Fe}{2} lines considered (see Figure \ref{11}). Note also that \citet{b9} used mainly measurements from lines from multiplets 37 and 38.

We found a correlation between total EW \ion{Fe}{2} and EW [\ion{O}{3}] $r= - 0.39$, $P < 0.0001$. The correlation coefficient was significantly lower for the lines obtained from I Zw 1 ($r = - 0.20$, $P = 0.0006$) than for the other three groups (see Table \ref{tbl-9}). We also analyzed relationships between EW \ion{Fe}{2} line groups and EW [\ion{O}{3}]/EW H$\beta$. For total EW \ion{Fe}{2} we found r= - 0.46, P$<$0.0001 (Figure \ref{17}, right), but for lines from I Zw 1 group r= - 0.28, P$<$0.0001.

No significant correlations were found between EW \ion{Fe}{2} and EW [\ion{N}{2}] lines (Table \ref{tbl-9}).

We also investigated whether there were any trends between equivalent widths of \ion{Fe}{2} $F$, $S$, $G$ line groups and equivalent widths of NLR, ILR and VBLR H$\alpha$ and H$\beta$ components (Table \ref{tbl-9}). We found no significant correlations.

 The correlations between EW \ion{Fe}{2} and the Doppler widths of H$\beta$ components are also considered (Table \ref{tbl-10}). As was expected, an inverse correlation was found between EW \ion{Fe}{2} and the widths of the broad H$\beta$ components (ILR and VBLR, as well as for the H$\beta$ FWHM, $r \sim  - 0.30$, $P < 0.0001$). These correlations are part of EV 1. But, contrary to this, we found a positive correlation ($r = 0.30$, $P < 0.0001$) between the EW \ion{Fe}{2} and the width of NLR component. 

\subsection{Relations among emission line strength and continuum luminosity}

 In Sec 3.3. we have confirmed the anti-correlation EW \ion{Fe}{2} vs. EW [\ion{O}{3}] (also EW \ion{Fe}{2} vs. EW [\ion{O}{3}]/EW H$\beta$). To try to understand the underlying physics which governs this anti-correlation, we examined its dependence of the continuum luminosity and redshift.

\begin{figure*}
\includegraphics[width=0.5\textwidth]{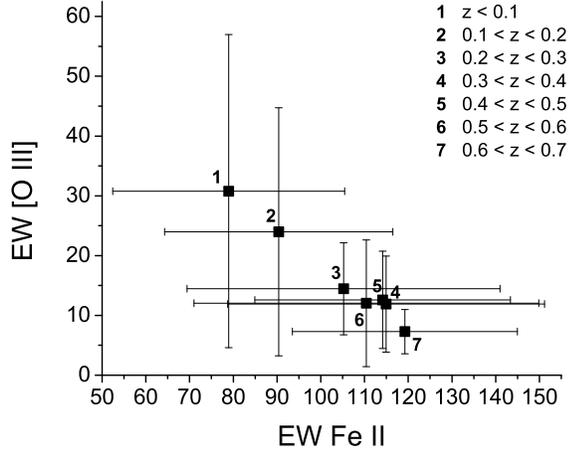}
\caption{Relationship between the equivalent widths, EW [\ion{O}{3}]$\lambda$ 5007 \AA \ and EW \ion{Fe}{2}, binned for cosmological redshift.  The error bars are the dispersions in each sub-sample within different redshift bins.}
\label{19}
\end{figure*}

Because of selection effects, cosmological redshift, $z$, and continuum luminosity are highly correlated in our sample, so it is difficult to clearly distinguish which influence dominates in some correlations. Since our sample has a more uniform redshift distribution than continuum luminosity distribution (see Figure \ref{3}), we binned equivalent widths of the [\ion{O}{3}] and \ion{Fe}{2} lines within the $z = 0$ -- 0.7 range, using a  $\Delta z = 0.1$ bin size. The binned data are presented in Figure \ref{19}.  It can be seen that as redshift increases in the $z < 0.4$ range, EW \ion{Fe}{2} also increases, but EW [\ion{O}{3}] decreases (Figure \ref{19}). For $z > 0.4$, the trend is not so obvious, probably due to the larger scatter of the data at higher redshifts.

\begin{figure*}
\includegraphics[width=0.40\textwidth]{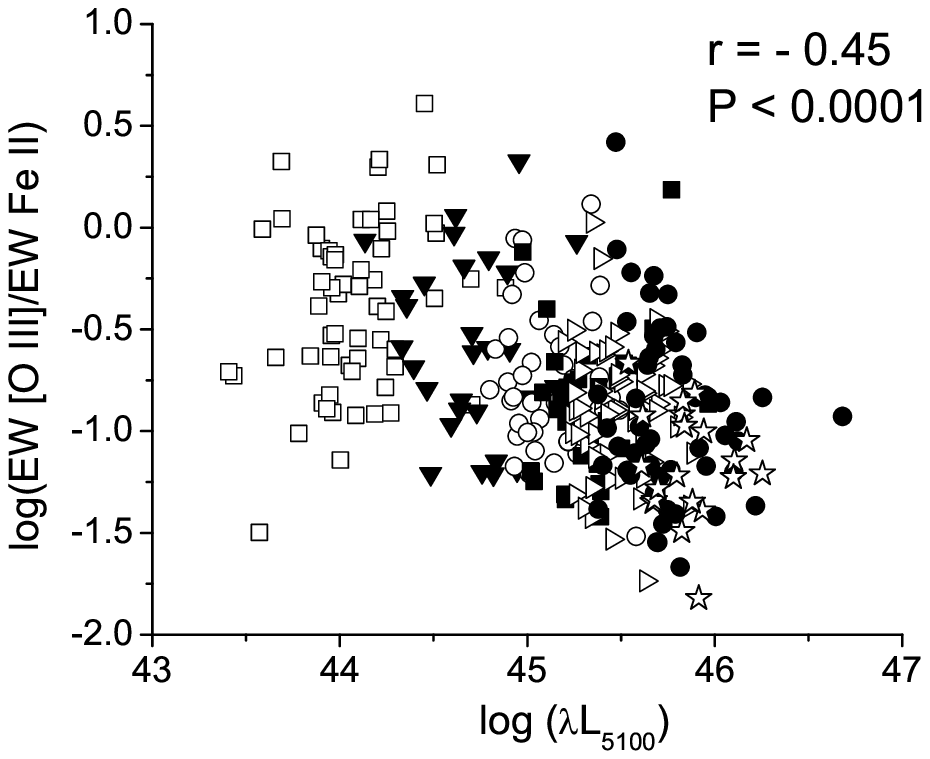}
\includegraphics[width=0.40\textwidth]{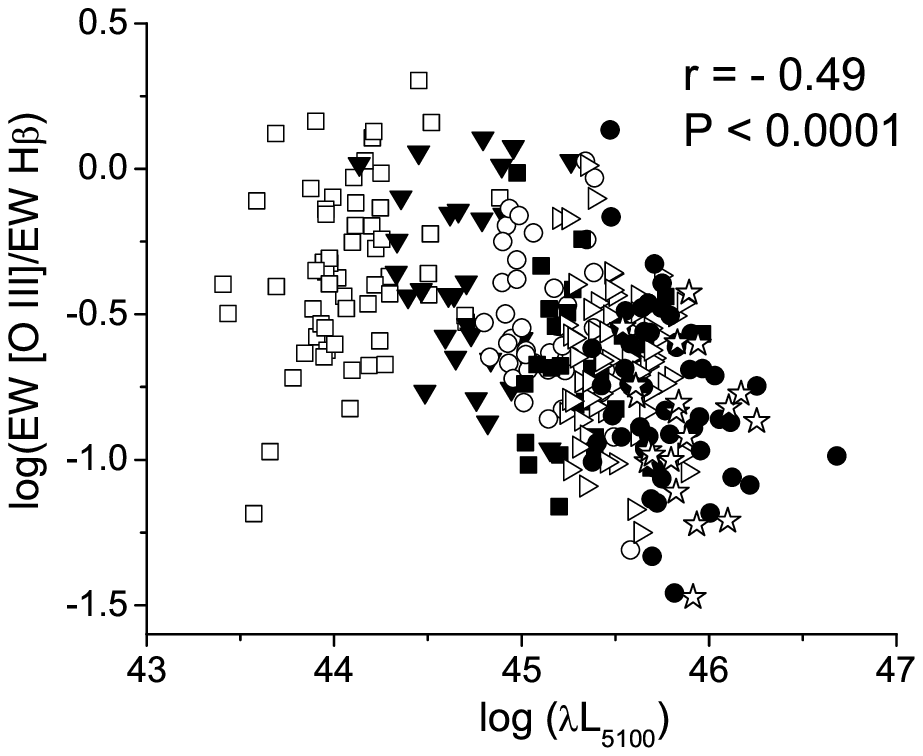}
\caption{First panel: EW [\ion{O}{3}]/EW \ion{Fe}{2} ratio vs. continuum luminosity ($\lambda L_{5100}$); second panel: the same for EW [\ion{O}{3}]/EW H$\beta$ ratio. Objects with redshift within range z$<$0.1 are denoted with open squares, 0.1$<$z$<$0.2 with filled triangles, 0.2$<$z$<$0.3 with open circles, 0.3$<$z$<$0.4 with filled squares, 0.4$<$z$<$0.5 with open triangles, 0.5$<$z$<$0.6 with filled circles and 0.6$<$z$<$0.7 with stars. [\ion{O}{3}] includes only the $\lambda$5007 \AA \ component.}
\label{20}
\end{figure*}

We derived correlations coefficients for the ratios of the [\ion{O}{3}], \ion{Fe}{2} and H$\beta$ lines and L$_{\lambda5100}$ (also with redshift, see Figure \ref{20}, Table \ref{tbl-11}). A significant anti-correlation was found for the ratio of EW [\ion{O}{3}]/EW \ion{Fe}{2} vs. L$_{\lambda5100}$ and for EW [\ion{O}{3}]/EW \ion{Fe}{2} vs. $z$ ($r= - 0.46$, $P = 0$ and $r= - 0.48$, $P = 0$, respectively). We found that the ratios of EW [\ion{O}{3}]/EW H$\beta$ and EW H$\beta$ NLR/EW \ion{Fe}{2} also decrease as L$_{\lambda5100}$ (redshift) increases. As in the previous case, the difference between the P--values for correlation with L$_{\lambda5100}$ and with $z$ is very small, so we cannot tell which effect dominates. In contrast to this, for EW H$\beta_{\mathrm{total}}$/EW \ion{Fe}{2} vs. L$_{\lambda5100}$, no significant trend is observed, but for EW H$\beta_{\mathrm{total}}$/EW \ion{Fe}{2} vs. $z$, there is a weak correlation ($r = 0.28$, $P =$ 1.3E-6).

Because of discrepancies in the properties observed between sub-samples within different H$\beta$ FWHM ranges (see Sec 3.1), we analyzed the relationships separately for the sub-samples with FWHM H$\beta<$3000 $\mathrm{kms^{-1}}$ and FWHM H$\beta>$3000 $\mathrm{kms^{-1}}$. It will be noticed that, in general, all the correlations considered are stronger for the FWHM H$\beta>$3000 $\mathrm{kms^{-1}}$ sub-sample than for spectra with narrower H$\beta$ lines (see Table \ref{tbl-11}). The only exception is the correlation of EW H$\beta$ NLR/EW \ion{Fe}{2} vs. L$_{\lambda5100}$, which is more significant for the FWHM H$\beta<$3000 $\mathrm{kms^{-1}}$ subsample.

It is interesting to connect these anti-correlations (EW \ion{Fe}{2} vs. EW [\ion{O}{3}] and considered ratios vs. L$_{\lambda5100}$) with the Baldwin effect. \citet{b22} confirmed a correlation between the Baldwin effect and some of the emission parameters which define Eigenvector 1 correlations (which are related to EW [\ion{O}{3}] - EW \ion{Fe}{2} anti-correlation). Also, it has been found that the [\ion{O}{3}] lines show strong Baldwin effect \citep{b36}. On the other hand, no Baldwin effect has been noticed for the optical \ion{Fe}{2} and H$\beta$ lines \citep{b36}, or even an inverse Baldwin Effect has been reported: \citet{b104} found an inverse Baldwin effect in H$\beta$ and \citet{b160} found one for both H$\beta$ and optical \ion{Fe}{2} lines.

Because of this, we examined the dependence of the equivalent widths of \ion{Fe}{2}, H$\beta$ and [\ion{O}{3}] lines vs. L$_{\lambda5100}$ and $z$ in our sample. Since  H$\beta$ is decomposed in NLR, ILR and VBLR and the iron lines are separated in the groups they can be considered in more detail. Also, we performed separate analyses for the sub-samples with different H$\beta$ FWHM ranges (Table \ref{tbl-11}).

\begin{figure*}
\includegraphics[width=0.40\textwidth]{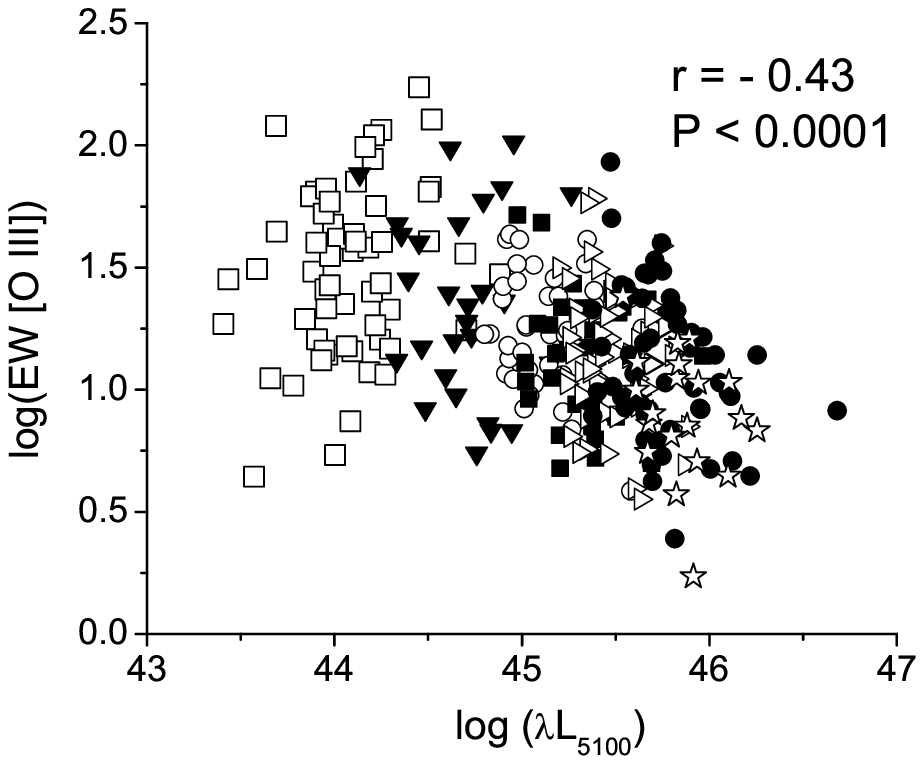}
\includegraphics[width=0.40\textwidth]{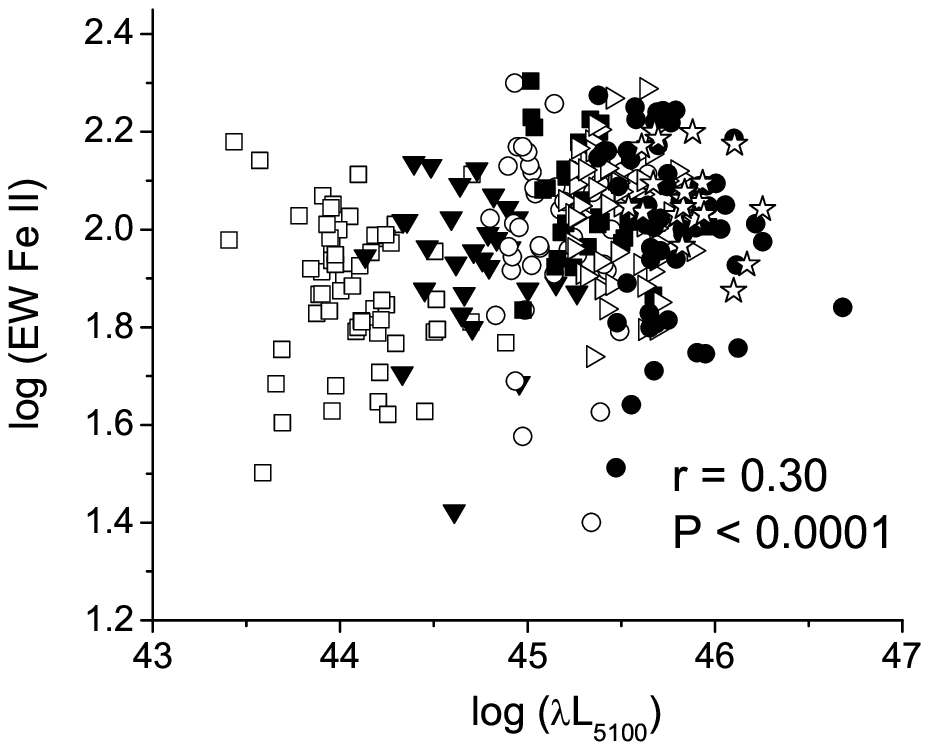}
\caption{The Baldwin effect is significant for the [\ion{O}{3}] lines (first panel), while an inverse Baldwin effect is detected for the optical \ion{Fe}{2} lines (second panel). Symbols have the same meaning as in Figure \ref{20}.}
\label{21}
\end{figure*}

Analyzing the total sample (i.e., the whole H$\beta$ FWHM range), it is obvious that \ion{Fe}{2} lines show an inverse Baldwin effect, which is specially strong for the central ($r = 0.33$, $P < 0.0001$) and red part ($r = 0.44$, $P < 0.0001$) of the \ion{Fe}{2} shelf ($S$ and $G$ groups), as well as for the group of lines from I Zw 1 ($r = 0.32$, $P < 0.0001$). The correlation of total \ion{Fe}{2} vs. L$_{\lambda5100}$ is presented in Figure \ref{21}. What is interesting is that a strong inverse Baldwin effect is not observed for the $F$ lines (blue part) ($r = 0.16$, $P = 0.004$).  In the relationship between H$\beta$ components and continuum luminosity for total sample, no trend is observed for broad H$\beta$ (ILR and VBLR), but it can be noticed that the narrow H$\beta$ component anti-correlates with continuum luminosity ($r= - 0.36$, $P < 0.0001$), -- i.e. they show a Baldwin effect -- as well as the [\ion{O}{3}] lines, for which the previously found trend is confirmed ($r= - 0.43$, $P < 0.0001$ -- see Figure \ref{21}).

 If we compare these correlations with the corresponding correlations with redshift, it is obvious that the iron lines (total \ion{Fe}{2} and the multiplet groups) correlate more strongly with redshift than with luminosity.  In the case of other lines we consider, the difference between the correlations with L$_{\lambda5100}$ and $z$, is not so significant. Generally, it seems that the lines which show an inverse Baldwin effect (\ion{Fe}{2} and H$\beta$ VBLR) correlate more significantly with redshift than with luminosity, while the lines which show a (normal) Baldwin effect ([\ion{O}{3}] and H$\beta$ NLR) have more significant correlations with continuum luminosity.

  We find that the coefficients of correlations between the EWs of lines and the continuum luminosity (and $z$) depend on the H$\beta$ FWHM range of the sub-sample. It seems that continuum luminosity affects iron lines and H$\beta$ more for the subsample with narrower H$\beta$ line (FWHM H$\beta<$3000 $\mathrm{kms^{-1}}$). In that sub-sample all \ion{Fe}{2} groups show a stronger inverse Baldwin effect, H$\beta$ NLR decreases more strongly with an increase in continuum luminosity, and an inverse Baldwin effect is also observed for H$\beta$ VBLR ($r = 0.41$, $P < 0.0001$). This is not seen in the FWHM H$\beta>$3000 $\mathrm{kms^{-1}}$ sub-sample. In contrast with this the Baldwin effect is stronger in the [\ion{O}{3}] lines for the FWHM H$\beta>$3000 $\mathrm{kms^{-1}}$ subsample.

\section{Discussion}

From our analysis of optical \ion{Fe}{2} lines ($\lambda\lambda$ 4400-5400 \AA), we can try to investigate physical and kinematical characteristics of the \ion{Fe}{2} emitting region in AGN as well as the connection between the \ion{Fe}{2} and other emission regions.  We have investigated above correlations between the \ion{Fe}{2} emission properties and some spectral features. Correlations of the \ion{Fe}{2} lines were considered separately for different multiplet groups, which enable more detailed analysis. Here we give some discussion of the results we have obtained.

\subsection{The \ion{Fe}{2} line group ratios -- possible physical conditions in the \ion{Fe}{2} emitting region}

Although we used very simple approximations to calculate the intensities of the observed \ion{Fe}{2} lines, we found that the calculated intensities can  satisfactorily fit the \ion{Fe}{2} shelf within $\lambda\lambda$ 4400-5500 \AA \ range. This approach is very simplified, but in some cases, it enables better fit than much more complicated theoretical models (see Appendix B). We included excitation temperature in calculation
and found that it is in the most cases around 10000 K (9646$\pm$2143 K), see Figure \ref{4}. Roughly estimated temperature of the \ion{Fe}{2} emitting region from our fitting (assumed template) is in a good agreement with the prediction earlier given in literature \citep[$\sim$7000 K,][]{b4}.

 We found that the most intensive emission in the optical part arises in transitions with the lower term b${\ }^4F$ (multiplets 37 and 38). We also considered the ratios of multiplet groups. The line ratios may indicate some physical properties.  For example, it has been shown that the Balmer line ratios are velocity dependent in AGN  \citep[see e.g.][]{sh1,sh2,cre,st90,st91,sne} and this is probably related to a range of physical conditions (electron temperature and density) and  to the radiative transfer effects \citep[see e.g.][]{pop03,pop06}. In general, the ratios between F, G and S groups indicate the ratio between the blue, red and central part of the \ion{Fe}{2} shelf around H$\beta$ lines. As can be seen in Table \ref{tbl-5}, there is correlation between $F/G$, $F/S$ and $G/S$ ratios vs. FWHM of H$\beta$ line. It is interesting that this correlation is not present  when we consider the cases where FWHM(H$\beta$)$<$ 3000 $\rm km s^{-1}$.  We should note here that for the case FWHM(H$\beta$)$<$ 3000 $\rm km s^{-1}$ we had only $\Delta$FWHM(H$\beta$)$\approx$2000 $\mathrm{km\ s^{-1}}$ and this may affect the obtained results. As we mentioned above, the separation of the sample into two sub-samples according to the H$\beta$ FWHM is similar as given in \citet[][and reference therein]{b0001}. It seems that the characteristics of the Pop A (FWHM H$\beta<$4000$\mathrm{kms^{-1}}$) and B (FWHM H$\beta>$4000$\mathrm{kms^{-1}}$) of AGN (as proposed by the above-mentioned authors) can be recognized in the \ion{Fe}{2} line ratios: namely, the correlations and trends observed in Table \ref{tbl-5} indicate that the ratios of blue and red (as well as blue and central) parts of \ion{Fe}{2} shelf, anti-correlate with continuum luminosity more strongly in Pop A than in Pop B. Also, for Pop B, all considered flux ratios (F/G, F/S and G/S) increase with increasing of H$\beta$ width and with decreasing \ion{Fe}{2} shift, while that trend is not observed in Pop A.

 The observed anti-correlations of $F/G$ and $F/S$ ratios vs. L$_{\lambda5100}$  could be affected by the differing degrees of inverse Baldwin effect observed for the \ion{Fe}{2} lines from the three line groups since the equivalent widths of the $S$ and $G$ groups increase more significantly with  $L_{\lambda5100}$ ($r = 0.33$ and $r = 0.43$,  $P < 0.0001$, respectively), than those of the $F$ group ($r = 0.16$, $P = 0.04$, see Table \ref{tbl-11}). Note here that the ratio of $G/S$ does not correlate with the continuum. Also, since the inverse Baldwin effect for the $S$  and $G$ groups is stronger for the FWHM H$\beta<$3000 $\mathrm{kms^{-1}}$ sub-sample, the correlations of $F/G$ and $F/S$ vs. L$_{\lambda5100}$ are stronger for FWHM H$\beta<$3000 $\mathrm{kms^{-1}}$ one (Table \ref{tbl-5}).

From the above-mentioned correlations, we can ask some intriguing questions about the physical conditions and processes in the \ion{Fe}{2} emitting region: (1) which processes cause the increase of \ion{Fe}{2} equivalent width with increasing L$_{\lambda5100}$? Note that the equivalent widths of the majority of other emission lines in AGN spectra decrease with increasing L$_{\lambda5100}$ \citep[see][]{b36}; (2) why is the inverse Baldwin effect correlation lower for the $F$ group comparing to the $S$ and $G$ groups? and (3) what causes that inverse Baldwin effect for \ion{Fe}{2} (in $S$ and $G$) to be more significant in the sub-sample with FWHM H$\beta<$3000 $\mathrm{kms^{-1}}$, than in one with broader H$\beta$ line? Answering some of these questions may lead to a better understanding of the complex \ion{Fe}{2} emitting region.

\subsection{Location of the \ion{Fe}{2} emitting region}

\citet{b001} and \citet{b31} noted earlier that \ion{Fe}{2} lines may originate in the intermediate-line region, which may be the transition from the torus to the BLR. Recently, other authors also found that the optical \ion{Fe}{2} emission may arise from a region in the outer parts of the BLR or several times larger than the H$\beta$ one \citep{b51,b151,b31,b66,b222,baa}.

To explore the kinematics of the \ion{Fe}{2} emitting region, which may indicate the location of the \ion{Fe}{2} emission, we compared the derived  widths and shifts of the \ion{Fe}{2} and the H$\beta$ and H$\alpha$ components. As can be
seen in Figure \ref{14} and Figure \ref{15}, there is an indication that \ion{Fe}{2} emitting region is located as the IL-emitting region. Moreover, the correlation between the Doppler widths of the \ion{Fe}{2} and IL regions is significantly higher than for the VBLR and
NLR. There is no significant correlation between the Doppler widths of the \ion{Fe}{2} and NL
regions ($r= 0.01$, $P=0.87$ for H$\alpha$, $r= 0.05$, $P=0.40$ for H$\beta$), but
there is a correlation between the \ion{Fe}{2} and VBLR regions ($r=0.66$, $P < 0.0001$ for H$\alpha$, $r = 0.45$, $P < 0.0001$ for H$\beta$). This may indicate that one more component of the \ion{Fe}{2} lines is arising from the
VBLR.  Note here that, due to the complex \ion{Fe}{2} template, we assumed only one component  for each \ion{Fe}{2} line (see Sect. 2.2). The reasonable fits of the \ion{Fe}{2} shelf indicate that,
the VBLR component of \ion{Fe}{2} should be significantly weaker
than the ILR one.  

\begin{figure*}

\includegraphics[width=0.37\textwidth]{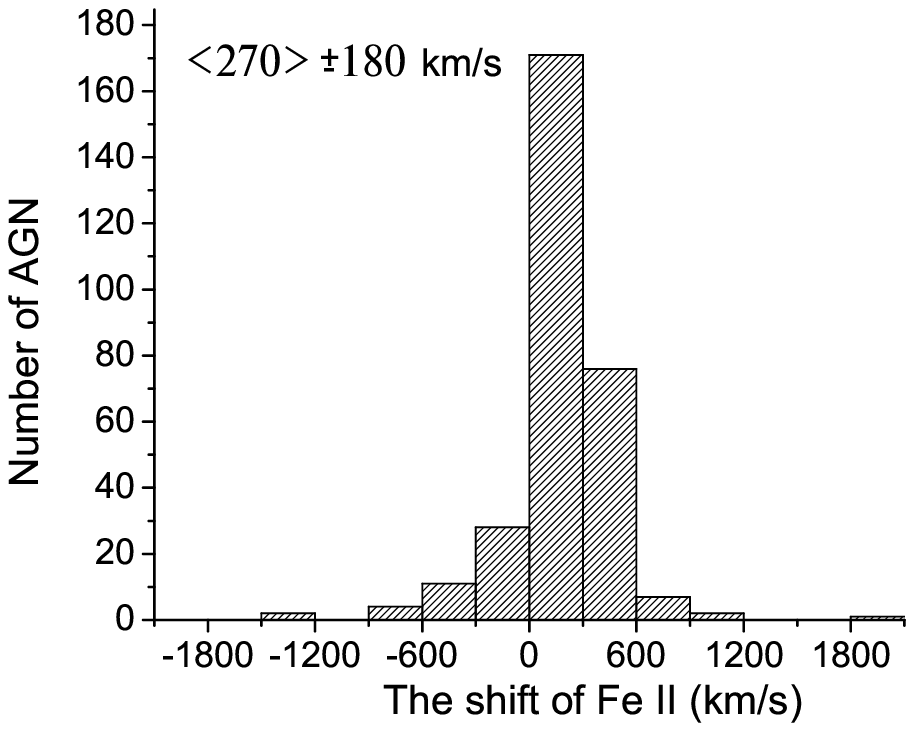}
\includegraphics[width=0.37\textwidth]{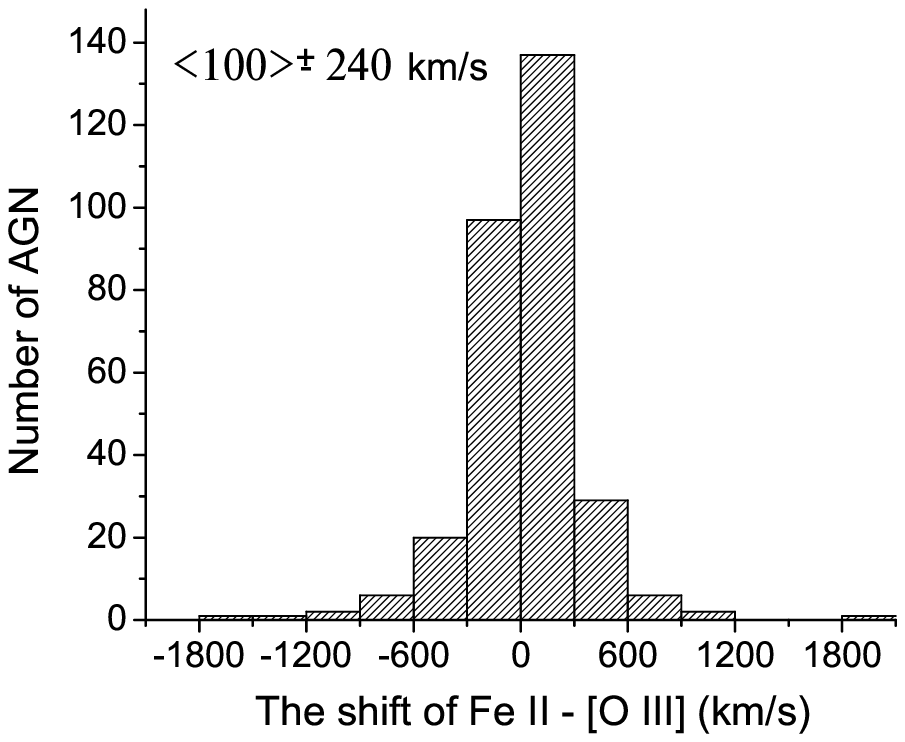}
\caption{Distribution of the \ion{Fe}{2} shift with respect to the transition wavelength (left) and distribution of the \ion{Fe}{2} shift with respect to the [\ion{O}{3}] lines (right).}
\label{18}
\end{figure*}

These indications are also supported by the correlation between the shift of the \ion{Fe}{2} lines and the ILR components of H$\alpha$ and H$\beta$.
The distribution of the \ion{Fe}{2} line shifts is shown in Figure \ref{18}. The shift is obtained with respect to the transition wavelength (Figure \ref{18}, left) and with respect to the narrow lines (Figure \ref{18}, right). We found that \ion{Fe}{2} lines are slightly redshifted, which is in agreement with the results of \citet{b51}. But in this case, the average value of the \ion{Fe}{2} shift  relative to the transition wavelength is $270\pm180$ $\mathrm{kms^{-1}}$, while the average value of the \ion{Fe}{2} shift with respect to the narrow lines is $100\pm240$ $\mathrm{kms^{-1}}$. This implies a significantly smaller redshift than found in \citet{b51}. A slightly redshifted \ion{Fe}{2} emission may indicate that \ion{Fe}{2} lines arise in an inflowing region \citep[see][]{ba}.

\subsection{EW \ion{Fe}{2} vs. EW [\ion{O}{3}] anti-correlation}

 One of the problems mentioned in the introduction is the anti-correlation
between the equivalent widths of the [\ion{O}{3}] and \ion{Fe}{2} lines which is related to EV1 in the analysis of \citet{b9}.
Some physical causes proposed to explain Eigenvector 1 correlations are: (a) Eddington ratio L/$\mathrm{L_{Edd}}$ \citep{b9,b22}, (b) black hole mass $M_{BH}$, and (c) inclination angle \citep{b25,b26,b27}.  \citet{be} also suggested that EV1 may be related to AGN evolution.

To try to understand the EW \ion{Fe}{2} vs. EW [\ion{O}{3}] anti-correlation, we examined its relationship to continuum luminosity and redshift (see Sec 3.4, Table \ref{tbl-11}). We found an anti-correlation between the EW [\ion{O}{3}]/EW \ion{Fe}{2} ratio and L$_{\lambda5100}$. Also, we examined the relations of equivalent widths of \ion{Fe}{2} and [\ion{O}{3}] lines vs. L$_{\lambda5100}$.  We confirmed a strong Baldwin effect for [\ion{O}{3}] lines, and an inverse Baldwin effect for EW \ion{Fe}{2} lines, i.e. we found that as continuum luminosity increases, EW \ion{Fe}{2} also increases, but EW [\ion{O}{3}] decreases. This implies that the EW \ion{Fe}{2} - EW [\ion{O}{3}] anti-correlation may be influenced by Baldwin effect for [\ion{O}{3}] and an inverse Baldwin effect for \ion{Fe}{2} lines. Also, in our analysis we found that the strength of the Baldwin effect depends on the H$\beta$ FWHM of the sample (see Table \ref{tbl-11}). Note that H$\beta$ FWHM is one of the parameters in Eigenvector 1. As it is mentioned in the introduction (Sec 1), some indications of connection between Baldwin effect and EV1 have been given in \citet{b22} and  \citet{bf}. \citet{Lu} found that the significance of the EV1 relationships is a strong function of continuum luminosity, i.e., the relationships between EW \ion{Fe}{2}, EW [\ion{O}{3}] and H$\beta$ FWHM can be detected only in a high-luminosity sample of AGNs (log$\lambda L_{5100}>44.7$), as well as a Baldwin effect for  [\ion{O}{3}] lines\footnote{ The majority of objects in our sample have continuum luminosities in the range $44.5<$log$\lambda L_{5100}<46$ range (see Fig \ref{3}).}

The origin of the Baldwin effect is still not understood and is a matter of debate.  The increase of the continuum luminosity may cause a decrease of the covering factor, or changes in the spectral energy distribution (softening of the ionizing continuum) which may result in the decrease of EWs. The inclination angle may also be related to Baldwin effect \citep[for review see][]{b163}. The physical properties which are usually considered as a primary cause of the Baldwin effect are: M$_{BH}$ \citep{b161,b162}, L/L$\odot$ \citep{b9,b22}, and changes in gas metalicity \citep{b36}. Also, a connection between Baldwin effect and AGN evolution is possible \citep[see][]{b163}.

We investigated if the correlations observed between L$_{\lambda5100}$ and the equivalent widths of the lines (as well as EW [\ion{O}{3}] vs. EW \ion{Fe}{2}) are primarily caused by evolution. Since continuum luminosity is strongly correlated with cosmological redshift in our sample, it is difficult to separate luminosity from evolutionary effect.  The exception are correlations of EW \ion{Fe}{2} vs. $z$, as well as the ratio EW H$\beta_{\mathrm{total}}$/EW \ion{Fe}{2} vs. $z$, which are significantly stronger than correlations of the same quantities with L$_{\lambda5100}$. This implies that inverse Baldwin effect of \ion{Fe}{2} may be governed first of all by an evolutionary effect. This result is in agreement with results of \citet{b163}.

\section{Conclusions}

In this paper we have investigated characteristic of the \ion{Fe}{2} emission region, using the sample of 302 AGN from SDSS.
 To analyze the \ion{Fe}{2} emission, an \ion{Fe}{2} template was constructed by grouping the strongest \ion{Fe}{2} multiplets into three groups, according to the atomic properties of transitions.

We have investigated the correlations of these \ion{Fe}{2} groups and their ratios with other lines in spectra. In this way, we tried to find some physical connection between the \ion{Fe}{2} and other emitting regions, as well as to connect \ion{Fe}{2} atomic structure with the physical properties of the emitting plasma.

We also investigated the kinematical connection between \ion{Fe}{2} and the Balmer emission region. In particular the anti-correlation of EW \ion{Fe}{2} -- EW [\ion{O}{3}] and its possible connection with AGN luminosity and evolution were analzed. From our investigation we can outline the following conclusions:

\begin{enumerate}

\item  We have proposed here an optical \ion{Fe}{2} template for the $\lambda\lambda$4400-5500 \AA \ range, which consists of three groups of  \ion{Fe}{2} multiplets, grouped according to the lower terms of transitions ($F$, $S$ and $G$), and an additional group of lines reconstructed from the I Zw 1 spectrum. We found that template can satisfactorily fit the \ion{Fe}{2} lines. In spectra in which \ion{Fe}{2} emission lines have different relative intensities than in I Zw 1, this template fit better than empirical and theoretical templates based on I Zw 1 spectrum (see more in Appendix B). Using this template, we are able to consider different groups of transitions which contribute to the blue, central and red part of \ion{Fe}{2} shelf around the H$\beta$+[\ion{O}{3}] lines.

\item We find that the ratios of different parts of the iron shelf ($F/G$, $F/S$, and $G/S$) depend of some spectral properties such as: continuum luminosity, H$\beta$ FWHM, shift of \ion{Fe}{2}, and $\mathrm{H\alpha}$/$\mathrm{H\beta}$ flux ratio. Also, it is noticed that spectra with H$\beta$ FWHM greater and less than $\sim$3000 $\mathrm{kms^{-1}}$ have different properties which is reflected in significantly different coefficients of correlation between the parameters.

\item We found that the \ion{Fe}{2} emission is mainly characterized with a random velocity of $\sim$1400 $\mathrm{kms^{-1}}$ that corresponds to the ILR origin, which is also supported by the significant correlation between the width of \ion{Fe}{2} and H$\alpha$, H$\beta$ ILR widths. This is in agreement with the previous investigations \citep{b31,b51,b151}, but unlike the earlier investigations, we found a slight correlation with the width of VBLR.  Therefore, it is possible that \ion{Fe}{2} is partly produced in the VBLR and cannot be resolved from continuum luminosity. Also, we found that the \ion{Fe}{2} lines  are slightly redshifted relative to the narrow lines ($\sim$ 100 $\mathrm{kms^{-1}}$) (see Figure \ref{18}).

\item  The Balmer lines were decomposed into NLR, ILR and VBLR components and relationships among H$\beta$ components and some spectral properties were investigated. We found a positive correlation between the width of narrow lines and EW \ion{Fe}{2} ($r=0.30$, $P < 0.0001$),  while the width of the broad H$\beta$ components anti-correlates with EW \ion{Fe}{2}. Also, we found a Baldwin effect in the case of H$\beta$ NLR component, while the ILR component showed no correlation with continuum luminosity, and VBLR component shows an inverse Baldwin effect for the FWHM $<$ 3000 $\mathrm{kms^{-1}}$ subsample.

\item  We confirm in our sample the anti-correlation between EW \ion{Fe}{2} and EW [\ion{O}{3}]  which is related to Eigenvector 1 (EV1) in \citet{b9} and we examined its dependence on the continuum luminosity and redshift. We found an inverse Baldwin effect for \ion{Fe}{2} lines from the central and red part of the \ion{Fe}{2} shelf ($S$ and $G$), but for \ion{Fe}{2} lines from blue part ($F$) no correlation was observed. A Baldwin effect was confirmed for the [\ion{O}{3}] lines. Since EW \ion{Fe}{2} increases, and EW [\ion{O}{3}] decreases with increases of continuum luminosity (and also the trend of decreasing of EW [\ion{O}{3}]/EW \ion{Fe}{2} ratio with luminosity is obvious), the observed EW \ion{Fe}{2} vs. EW [\ion{O}{3}] anti-correlation is probably due to the same physical reason which causes the Baldwin effect.  Moreover, it is observed that the coefficients of correlation due to Baldwin effect depend on H$\beta$ FWHM range of a sub-sample, which also implies the connection between the Baldwin effect and EV1. A decreasing trend is observed for EW [\ion{O}{3}]/EW H$\beta$ vs. continuum luminosity.

\item  We found that the equivalent width of the \ion{Fe}{2} lines, as well as the EW H$\beta_{\mathrm{total}}$/EW \ion{Fe}{2} ratio have a stronger correlation with redshift, than with continuum luminosity. This implies that the inverse Baldwin effect of \ion{Fe}{2} may be primarily caused by evolution.

\end{enumerate}

\acknowledgments

This work is a part of the projects (146002): "Astrophysical Spectroscopy of Extragalactic Objects" and (146001):  "Influence of collisions with charged particles on astrophysical spectra", supported by Serbian Ministry of Science and Technological Development.
Data for the present study have been entirely collected at the SDSS
database. Funding for the SDSS and SDSS-II has been provided by the Alfred P.
Sloan Foundation, the Participating Institutions, the National Science
Foundation, the U.S. Department of Energy, the National Aeronautics and Space Administration,
the Japanese Monbukagakusho, the Max Planck Society, and the Higher Education Funding
Council for England. The SDSS Web Site is http://www.sdss.org/.
The SDSS is managed by the Astrophysical Research Consortium for the
Participating Institutions. The Participating Institutions are the American Museum of
Natural History, Astrophysical Institute Potsdam, University of Basel, University of
Cambridge, Case Western Reserve University, University of Chicago, Drexel
University, Fermilab, the Institute for Advanced Study, the Japan Participation Group, John Hopkins
University, the Joint Institute for Nuclear Astrophysics, the Kavli Institute for Particle
Astrophysics and Cosmology, the Korean Scientist Group, the Chinese Academy of Sciences
(LAMOST), Los Alamos National Laboratory, the Max-Planck-Institute for Astronomy (MPIA),
the  Max-Planck-Institute for Astrophysics (MPA), New Mexico State
University, Ohio State University,  University of Pittsburgh, University of Portsmouth,
Princeton University, the United States Naval Observatory, and the University of Washington. This research
has made use of NASA's Astrophysics Data System.



{\it Facilities:} \facility{Nickel}, \facility{HST (STIS)}, \facility{CXO (ASIS)}.



 We would like to thank M. Gaskell for very useful comments and suggestions.

\appendix
\section{The sample selection}

\begin{figure*}
\includegraphics[width=0.4\textwidth]{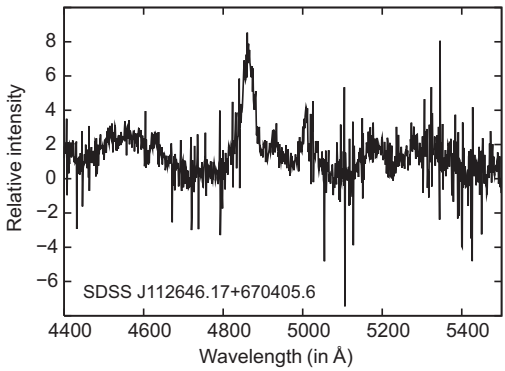}
\caption{Example of a spectrum rejected because of strong noise which inhibited a precise fit to the iron lines.}
\label{sum}
\end{figure*}

Using an SQL search with requirements mentioned in Sec 2.1, we obtained 497 AGN spectra with broad emission lines. From that number, 188 spectra were rejected because of noise in the iron emission line region. An example of rejected spectrum is shown in Figure \ref{sum}. Also, three spectra were rejected because of bad pixels, and one because of very strong double-peak emission in H$\beta$ (SDSS J154213.91$+$183500.0).

\begin{figure*}
\includegraphics[width=0.4\textwidth]{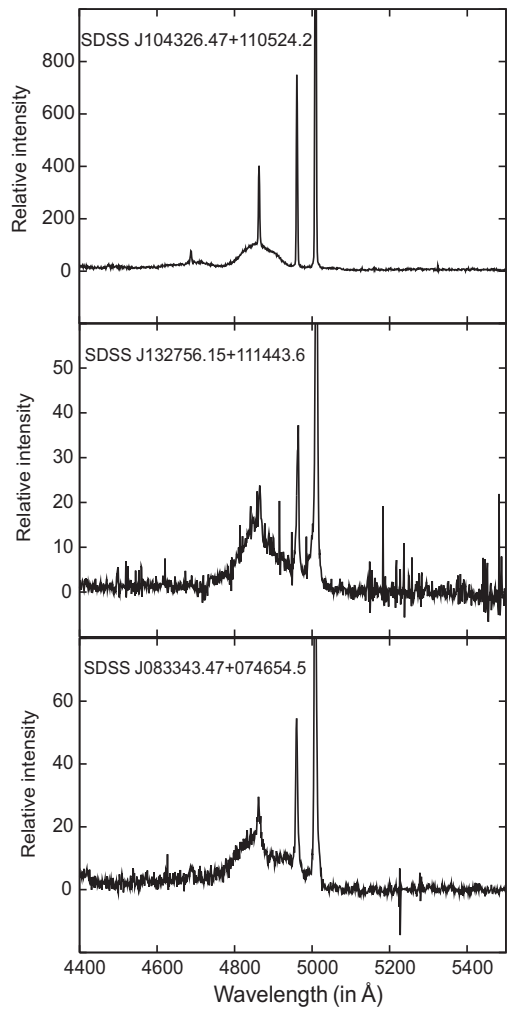}
\caption{The three spectra with broad H$\beta$ line and without \ion{Fe}{2} emission.}
\label{nemaFe}
\end{figure*}

 Three more spectra with broad H$\beta$ and practically without the \ion{Fe}{2} emission were  rejected since it was not possible to fit the \ion{Fe}{2} lines (SDSS J132756.15$+$111443.6, SDSS J104326.47$+$110524.2 and SDSS J083343.47$+$074654.5). These three spectra are shown in Figure \ref{nemaFe}. There is also a possibility that in these three objects the \ion{Fe}{2} lines are very broad and weak so that they cannot be distinguished from the continuum emission. 

The sample contains 302 AGN spectra (table with the SDSS identification and obtained parameters from the best fit is available electronically only, as an additional file at http://www.aob.rs/$\sim$jkovacevic/table.txt).

 To test the host galaxy contribution in the selected sample of 302 AGNs, we have compared our sample with the set of AGNs which host galaxy fraction is determined in \citet{VB}. We found 106 common objects. In the case of 48 AGNs (45\% of the sample) there is practically no host galaxy contribution (0\%), and in the rest of the objects, the contribution of the host galaxy is mainly smaller than 20\%.

\section{Comparison with other \ion{Fe}{2} templates}

We compared two \ion{Fe}{2} templates (one empirical and one theoretical) with our template. The empirical \ion{Fe}{2} template was taken from \citet{b302} using 46 broad and 95 narrow lines identified by \citet{b303} in the I Zw 1 spectrum within a $\lambda\lambda$4400-5500 \AA \ range. The \ion{Fe}{2} lines were fit with 6 free parameters (the width, shift and intensity for narrow and the width, shift and intensity for broad lines). We also considered the theoretical model from \citet{b304} calculated for log[$\mathrm{n_H}$/(cm$^{-3}$)]=11, [$\xi$]/(1 km s$^{-1}$)=20 and log[$\mathrm{\Phi_H}$/(cm$^{-2}$ s$^{-1}$)]=21, since we found that this gave the best fit for those values of physical parameters. With this model, \ion{Fe}{2} lines were fit with 3 free parameters (width, shift and intensity).

Figure \ref{5} shows the spectrum of SDSS J020039.15-084554.9, an object in which the iron lines have similar relative intensities to I Zw 1. We fit that spectrum with our template (a), with a template based on line intensities from I Zw 1 \citep{b302} (b) and with a template calculated by CLOUDY code \citep{b304} (c). We found that for this object all three templates fit the iron lines well.

In the case of the SDSS J$141755.54+431155.8$, however (Figure \ref{6}), the iron lines in the blue bump are slightly stronger compared to those in the red, and our model fit iron emission slightly better than the other two templates. The discrepancy between the blue and red \ion{Fe}{2} bumps is especially emphasized in the spectrum of the object SDSS J$111603.13+020852.2$ (Figure \ref{7}), which makes a significant difference for comparison with the \ion{Fe}{2} properties of I Zw 1 object. In this case, our template shows a small disagreement for the lines whose relative intensity is taken from I Zw 1, but other two models, cannot fit well this type of the \ion{Fe}{2} emission.  Disagreement is particularly strong in the blue bump (i.e., in the multiplets 37 and 38 of the $F$ group).  

\section{The relative intensities of \ion{Fe}{2} lines}

To estimate the relative intensities of \ion{Fe}{2} lines within the three groups,
we stated that
the intensity of a line ($m\to n$) should be proportional to the number
of emitters ($N_{m}$)\citep{b61,b611}, i.e. for optically
thin plasma (for \ion{Fe}{2} lines) one can write:
$I_{mn}=const\cdot\frac{A_{mn}}{\lambda_{mn}} \int_0^\ell N_m dx$,
where $A_{mn}$ is the probability of the transition, $\lambda_{mn}$ is the
transition wavelength, and $\ell$ is the depth of the
emitting region.

Of course, the density of emitters can be non-uniform across the region,
but we can approximate here that it is uniform, and
in the case of non-thermodynamical equilibrium it could be written \citep{b61}:
$$N_{m}\sim b(T,N_e) g_m\exp{(-E_{mn}/kT)},$$
where b(T,$N_e$) represents deviation from thermodynamical equilibrium. In
the case of lines which have the same lower level, one may
approximate the ratio as \citep{b611}:

\begin{equation}
\frac{I_1}{I_2}=\frac{b_1(T,N_e)}{b_2(T,N_e)}{(\frac{\lambda_2}
{\lambda_1})}^3\frac{f_1}{f_2}\cdot\frac{g_1}{g_2}\cdot e^{-(E1-E2)/kT}
\end{equation}
Assuming that $\frac{b_1(T,N_e)}{b_2(T,N_e)}\approx 1$ we obtained Eq. (1) for estimation of line ratios within one group.




\clearpage


\clearpage

\begin{figure*}

\includegraphics[width=0.60\textwidth]{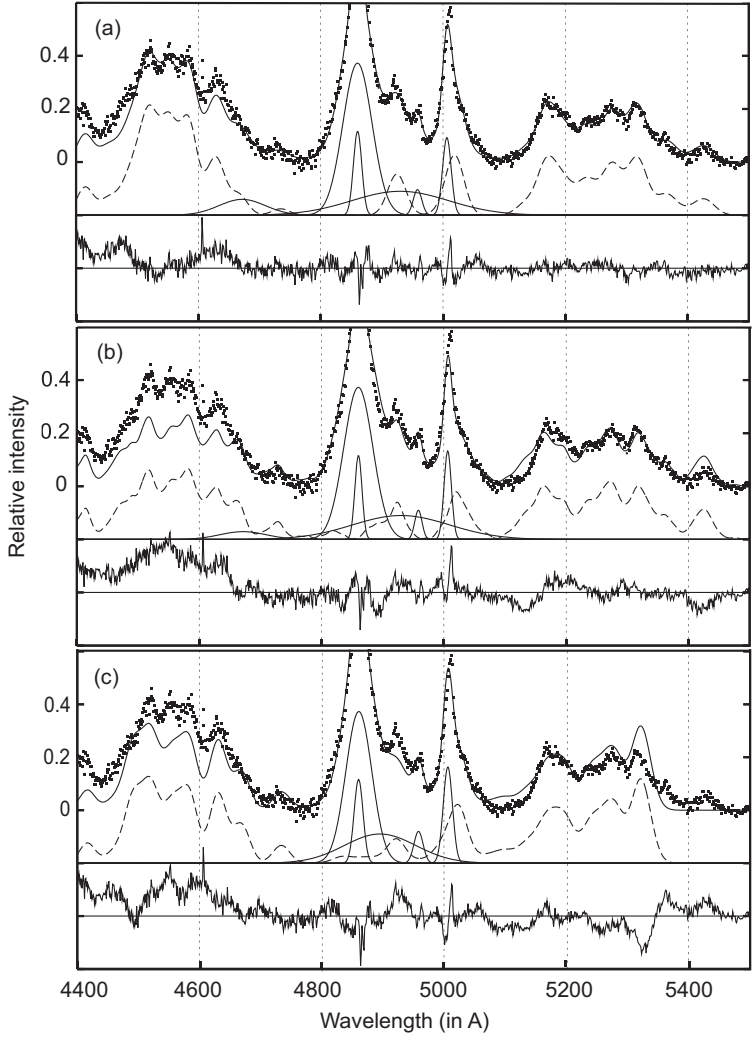}

\caption{Examples of fits to SDSS J$111603.13+020852.2$: with our template (a), with the empirical template of \citet{b302} (b) and with the theoretical template of \citet{b304} (c). In this object iron emission is much stronger in blue than in red bump. Our template show disagreement for  lines whose relative intensity is taken from I Zw 1, but other the three wavelength regions based on our three line groups fit the observed \ion{Fe}{2} well. The other two models cannot fit this kind of \ion{Fe}{2} emission well. }
\label{7}
\end{figure*}

\clearpage
\begin{table*}
\footnotesize
\begin{center}
\caption{A list of the 35 strongest \ion{Fe}{2} lines within the $\lambda\lambda$ 4400-5500 \AA \ region used to calculate the \ion{Fe}{2} template. In the first column are wavelengths (in air),  in the second multiplet numbers \citep{b105}, in the third are  terms of transitions, in the fourth are $gf$ values used for the template calculation, in the fifth are the references for the source of oscillator strengths, and in the 6th-8th columns are relative intensities, calculated for T=5000 K, 10000 K and 15000 K. Intensities of lines from the $F$, $S$ and $G$ groups are normalized to intensities of the $\lambda$4549.474 \AA, $\lambda$5018.44 \AA, \ and $\lambda$5316.6 \AA \ lines (respectively).\label{tbl-1}}
\vspace*{0.3cm}
\begin{tabular}{cccccccc}
\tableline\tableline
\vspace*{0.2cm}

Wavelength &Multiplet& Transition & gf & ref. &\multicolumn{3}{c} {Relative intensity}\\
 & &  &  &  & T=5000 K & T=10000 K & T=15000 K\\

\tableline

4472.929&37 &b${\ }^4F_{5/2}$ - z${\ }^4F^o_{3/2}$&4.02E-04&1  & 0.033   & 0.036& 0.037      \\
4489.183&37 &b${\ }^4F_{7/2}$ - z${\ }^4F^o_{5/2}$ &1.20E-03&1 & 0.105   & 0.110& 0.111      \\
4491.405&37 &b${\ }^4F_{3/2}$ - z${\ }^4F^o_{3/2}$&2.76E-03 &1 & 0.226   & 0.243& 0.249     \\
4508.288&38 &b${\ }^4F_{3/2}$ - z${\ }^4D^o_{1/2}$  &4.16E-03&2 & 0.345   & 0.367& 0.375      \\
4515.339&37 &b${\ }^4F_{5/2}$ - z${\ }^4F^o_{5/2}$  &3.89E-03&2& 0.333   & 0.348& 0.353      \\
4520.224&37 &b${\ }^4F_{9/2}$ - z${\ }^4F^o_{7/2}$  &2.50E-03&3 & 0.235   & 0.233& 0.233      \\
4522.634&38 &b${\ }^4F_{5/2}$ - z${\ }^4D^o_{3/2}$ &9.60E-03&3 & 0.827   & 0.859& 0.870      \\
4534.168&37 &b${\ }^4F_{3/2}$ - z${\ }^4F^o_{5/2}$  &3.32E-04&3 & 0.028   & 0.029& 0.030      \\
4541.524&38 &b${\ }^4F_{3/2}$ - z${\ }^4D^o_{3/2}$  &8.80E-04 &3 & 0.075   & 0.078& 0.079      \\
4549.474&38 &b${\ }^4F_{7/2}$ - z${\ }^4D^o_{5/2}$ &1.10E-02& 4& 1.000   & 1.000& 1.000    \\
4555.893&37 &b${\ }^4F_{7/2}$ - z${\ }^4F^o_{7/2}$&5.20E-03 &3 & 0.477   & 0.474& 0.474      \\
4576.340&38 &b${\ }^4F_{5/2}$ - z${\ }^4D^o_{5/2}$ &1.51E-03 &4& 0.136   & 0.136& 0.136      \\
4582.835&37 &b${\ }^4F_{5/2}$ - z${\ }^4F^o_{7/2}$ &7.80E-04  &3& 0.070   & 0.070& 0.070      \\
4583.837&38 &b${\ }^4F_{9/2}$ - z${\ }^4D^o_{7/2}$  &1.44E-02&2 & 1.420   & 1.353& 1.331      \\
4620.521&38 &b${\ }^4F_{7/2}$ - z${\ }^4D^o_{7/2}$  &8.32E-04&4 & 0.080   & 0.076& 0.075      \\
4629.339&37 & b${\ }^4F_{9/2}$ - z${\ }^4F^o_{9/2}$&4.90E-03&4 & 0.497   & 0.459& 0.447      \\
4666.758&37 &b${\ }^4F_{7/2}$ - z${\ }^4F^o_{9/2}$  &6.02E-04 &4 & 0.060   & 0.055& 0.054      \\
4993.358&36 &b${\ }^4F_{9/2}$ - z${\ }^6P^o_{7/2}$  &3.26E-04 &4 & 0.041   & 0.030& 0.027     \\                              5146.127&35 &b${\ }^4F_{7/2}$ - z${\ }^6F^o_{7/2}$  &1.22E-04&5  & 0.016   &0.011& 0.010        \\
\tableline                                                                                                              4731.453&43  &a${\ }^6S_{5/2}$ - z${\ }^4D^o_{7/2}$  &1.20E-03  &2 &0.025 &0.030& 0.032 \\
4923.927&42  &a${\ }^6S_{5/2}$ - z${\ }^6P^o_{3/2}$&2.75E-02 &4   &0.656 &0.693& 0.706 \\                                       5018.440&42  &a${\ }^6S_{5/2}$ - z${\ }^6P^o_{5/2}$  &3.98E-02&4   &1.000 &1.000& 1.000 \\                                    5169.033& 42 &a${\ }^6S_{5/2}$ - z${\ }^6P^o_{7/2}$ &3.42E-02 &1  &0.929 &0.854& 0.831 \\                                  5284.109& 41 &a${\ }^6S_{5/2}$ - z${\ }^6F^o_{7/2}$  &7.56E-04  &2 &0.022 &0.019& 0.018 \\
\tableline
5197.577&49 & a${\ }^4G_{5/2}$ - z${\ }^4F^o_{3/2}$ &7.92E-03 & 5 & 0.532  &0.620&0.652  \\
5234.625&49 & a${\ }^4G_{7/2}$ - z${\ }^4F^o_{5/2}$&8.80E-03 & 3 & 0.615  &0.695&0.723  \\
5264.812&48 & a${\ }^4G_{5/2}$ - z${\ }^4D^o_{3/2}$ &1.08E-03&1 &  0.075  &0.084&0.087  \\
5276.002&49 & a${\ }^4G_{9/2}$ - z${\ }^4F^o_{7/2}$ &1.148e-02 &2 & 0.861  &0.928&0.951  \\
5316.615&49 & a${\ }^4G_{11/2}$ - z${\ }^4F^o_{9/2}$&1.17E-02& 4 & 1.000  &1.000&1.000  \\
5316.784&48 & a${\ }^4G_{7/2}$ - z${\ }^4D^o_{5/2}$&1.23E-03   & 5& 0.089  &0.097&0.099  \\
5325.553&49 & a${\ }^4G_{7/2}$ - z${\ }^4F^o_{7/2}$&6.02E-04& 4   & 0.044  &0.047&0.048  \\
5337.732&48 & a${\ }^4G_{5/2}$ - z${\ }^4D^o_{5/2}$&1.28E-04  & 5 & 0.009  &0.010&0.010  \\
5362.869&48 & a${\ }^4G_{9/2}$ - z${\ }^4D^o_{7/2}$&1.82E-03  & 5 & 0.142  &0.146&0.148  \\
5414.073&48 & a${\ }^4G_{7/2}$ - z${\ }^4D^o_{7/2}$&1.60E-04  & 5 & 0.012  &0.012&0.013  \\
5425.257&49 & a${\ }^4G_{9/2}$ - z${\ }^4F^o_{9/2}$&4.36E-04&5    & 0.035  &0.035&0.035  \\
\tableline
 \tablerefs{(1) \citet{b100}, (2) \citet{b105}, (3) NIST Atomic Spectra Database (http://physics.nist.gov/PhysRefData/ASD/), (4) \citet{b101} and (5) http://www.pmp.uni-hannover.de/cgi-bin/ssi/test/kurucz/sekur.html}

\end{tabular}
\end{center}
\end{table*}

\begin{table*}
\footnotesize
\begin{center}
\caption{List of the lines taken from Kurucz database (http://www.pmp.uni-hannover.de/cgi-bin/ssi/test/kurucz/sekur.html). In first column are  wavelengths (in air), in the second oscillator strengths, and in the third relative intensities measured in I Zw 1.\label{tbl-2}}
\vspace*{0.3cm}
\begin{tabular}{c c c c}
\tableline\tableline
\vspace*{0.2cm}
Wavelength&Transitions&gf&Relative intensity\\
\tableline
4418.957&y${\ }^4G_{5/2}$ - e${\ }^4F_{3/2}$ &1.45E-02   & 3.00  \\
4449.616&y${\ }^4G_{9/2}$ - e${\ }^4F_{7/2}$ &2.58E-02   & 1.50  \\
4471.273&y${\ }^4D_{5/2}$ - f${\ }^4D_{3/2}$ &6.40E-03   & 1.20  \\
4493.529&y${\ }^4G_{10/2}$ - e${\ }^4F_{9/2}$ &3.74E-02  & 1.60 \\
4614.551&y${\ }^6P_{5/2}$ - ${\ }^6D_{7/2}$ &2.69E-03   &  0.70   \\
4625.481&x${\ }^4D_{1/2}$ - f${\ }^4D_{1/2}$ &8.51E-03    &0.70 \\
4628.786&x${\ }^4D_{5/2}$ - f${\ }^4D_{5/2}$ &1.83E-02   & 1.20 \\
4631.873&x${\ }^4D_{3/2}$ - f${\ }^4D_{3/2}$ & 1.34E-02   &0.60 \\
4660.593&$d^5 4s^2 {\ }^2G_{7/2}$ - $(4G)sp {\ }^2G_{7/2}$ &1.15E-03   & 1.00  \\
4668.923&x${\ }^4D_{5/2}$ - f${\ }^4D_{7/2}$ &3.13E-03    & 0.90   \\
4740.828&y${\ }^6P_{5/2}$ - e${\ }^6F_{7/2}$ &1.98E-03   &  0.50 \\
5131.210 &y${\ }^4P_{5/2}$ - e${\ }^4D_{5/2}$ &2.13E-03   & 1.1 \\
5369.190&e${\ }^4D_{5/2}$ - ${\ }^4D_{5/2}$ &1.23E-03  &    1.45 \\
5396.232&x${\ }^4F_{5/2}$ - f${\ }^4D_{3/2}$ &2.80E-03   &  0.40 \\
5427.826&b${\ }^4G_{11/2}$ - w${\ }^4F_{9/2}$ &2.17E-02   & 1.40\\
\tableline
\end{tabular}
\end{center}
\end{table*}

\begin{table*}
\small
\begin{center}
\caption{Correlations between widths, shifts and luminosities of the NLR, ILR and VBLR components of the H$\alpha$ and H$\beta$ lines. The values in brackets are for the sample without the 16 objects with highly redshifted H$\beta$ VBLR components ($>$4000 $\mathrm{kms^{-1}}$). The coefficient of correlation ($r$) is in the second column, the $P$-value which is a measure of significance of the correlation is given in the third column, and in the next columns are the coefficients $A$ and $B$ from the fit with a linear function: $H\beta=A+B\cdot H\alpha$.  The slopes are from regressions where $X$ is taken as the independent variable.
\label{tbl-3}}
\vspace*{0.3cm}
\begin{tabular}{l c c c c}
\tableline\tableline
\vspace*{-0.2cm}
\tiny{}&\tiny{} &\tiny{}&\tiny{}&\tiny{} \\
\vspace*{-0.2cm}
correlation (H$\alpha$ and H$\beta$) &r&P&A&B\\
\tiny{}&\tiny{} &\tiny{}&\tiny{}&\tiny{} \\
\tableline

H$\alpha$ NLR width vs.  H$\beta$ NLR width&  0.99 & $<0.0001$&$6.19\pm2.81$ & $0.97\pm0.01$ \\
H$\alpha$ ILR width vs.  H$\beta$ ILR width& 0.91&$<0.0001$  &$165.37\pm43.29$&$0.74\pm0.03$\\
H$\alpha$ VBLR width vs. H$\beta$ VBLR width &  0.50 (0.42)& $<0.0001$ &$1863.15\pm352.50$&$0.54\pm0.08$\\
\tableline

H$\alpha$ NLR shift  vs.  H$\beta$  NLR shift  &  0.97 & $<0.0001$&$16.40\pm3.58$ & $0.91\pm0.02$ \\
H$\alpha$ ILR shift  vs.  H$\beta$  ILR shift  & 0.90&$<0.0001$  &$-1.86\pm16.97$&$1.09\pm0.05$\\
H$\alpha$ VBLR shift vs. H$\beta$  VBLR shift &  0.18 (0.41)&0.041 ($<0.0001$)&$-93.64\pm45.30$&$0.06\pm0.03$ \\

\tableline
L H$\alpha$ NLR  vs. L H$\beta$ NLR & 0.93&$<0.0001$&$3.40\pm1.29$&$0.93\pm0.03 $ \\
L H$\alpha$ ILR  vs. L H$\beta$ ILR &0.96&$<0.0001$&$1.93\pm1.08$& $0.97\pm0.03$\\
L H$\alpha$ VBLR vs. L H$\beta$ VBLR&0.96&$<0.0001$&$0.70\pm1.03$& $0.99\pm0.02$ \\
\tableline

\end{tabular}
\end{center}
\end{table*}

\begin{table*}
\small
\begin{center}
\caption{Correlatios between equivalent widths (EW) of the \ion{Fe}{2} $F$, $S$ and $G$ multiplet groups. Relationships are fitted with function $Y=A+B\cdot X$. The coefficient of correlations, $r$, the corresponding measure of the significance of correlations, $P$, as well as the $A$ and $B$ coefficients, are shown in the table.\label{tbl-4}}
\vspace*{0.3cm}
\begin{tabular}{llcccc}
\tableline\tableline
\vspace*{-0.2cm}
\tiny{}&\tiny{} &\tiny{}&\tiny{}&\tiny{} &\tiny{}\\
\vspace*{-0.2cm}

{\rm X}&{\rm Y}&{\rm  r}&P&A& B\\
\tiny{}&\tiny{} &\tiny{}&\tiny{}&\tiny{} &\tiny{}\\
\tableline
EW Fe II F & EW Fe II S&0.76&$<0.0001$ &$0.21\pm1.00$& $0.66\pm0.03$\\
EW Fe II F&EW Fe II G& 0.80 &$<0.0001$ &$-0.17\pm1.29$& $0.96\pm0.04$\\
EW Fe II S &EW Fe II G&0.83&$<0.0001$ &$5.39\pm0.95$& $0.14\pm0.04$ \\
\tableline
\end{tabular}
\end{center}
\end{table*}

\begin{table*}
\small
\begin{center}
\caption{Correlations between the ratio of \ion{Fe}{2} line groups and spectral properties. In the first column is FWHM of broad H$\beta$ (ILR+VBLR), in the second and third columns are the Doppler widths of ILR and VBLR H$\beta$ components, in the fourth and fifth the Doppler width and shift of \ion{Fe}{2} lines, and in the eighth column is the continuum luminosity at $\lambda$5100 \AA. Correlation coefficients are calculated for total sample, for objects with H$\beta$ FWHM $<$ 3000 $\mathrm{kms^{-1}}$ (1), and with H$\beta$ FWHM $>$ 3000 $\mathrm{kms^{-1}}$ (2).  The correlations with $P<0.0001$ are printed in bold print.\label{tbl-5}}
\vspace*{0.3cm}
\begin{tabular}{c c c c c c c c}
\tableline\tableline
\vspace*{-0.2cm}
\tiny{}&\tiny{} &\tiny{}&\tiny{}&\tiny{} &\tiny{}&\tiny{} &\tiny{}\\
\vspace*{-0.2cm}
 & & H$\beta$ FWHM & w H$\beta$ ILR & w H$\beta$ VBLR &
 w Fe II & sh Fe II &
log $\lambda L_{5100}$\\

\tiny{}&\tiny{} &\tiny{}&\tiny{}&\tiny{} &\tiny{}&\tiny{} &\tiny{}\\
\tableline
F/G&r &{\bf 0.28}&{\bf 0.24}&{\bf 0.29}&0.18&{\bf - 0.27}&{\bf - 0.32}\\
&P &{\bf 7.13E-7}&{\bf 2.49E-5}&{\bf 2.37E-7}&0.002&{\bf 2.75E-6}&{\bf 1.82E-8}\\
 \tableline
(1) F/G&r &- 0.10&- 0.02&0.16&0.001&- 0.10&{\bf - 0.51}\\
&P         &0.20&0.831&0.046&0.987&0.189&{\bf 5.72E-12}\\
 \tableline
(2) F/G&r &{\bf 0.36}&0.27&0.31&0.17&- 0.30&{\bf- 0.36}\\
&P &{\bf 1.22E-5}&0.001&1.82E-4 &0.043&3.31E-4&{\bf 8.98E-6}\\
 \tableline
F/S&r &{\bf 0.41}&{\bf 0.26}&0.16&0.20&{\bf - 0.27}&- 0.15\\
&P &{\bf 1.95E-13}&{\bf 4.05E-6}&0.004&5.47E-4&{\bf 2.44E-6}&0.011\\
 \tableline
(1) F/S&r &- 0.21&- 0.18&0.07&-0.23&- 0.11&{\bf - 0.41}\\
&P &0.009&0.027&0.407&0.004&0.182&{\bf 7.90E-8}\\
 \tableline
(2) F/S &r &{\bf 0.59}&{\bf 0.33}&0.12&0.25&- 0.29&- 0.18\\
&P         &{\bf 1.35E-14}&{\bf 5.08E-5}&0.157&0.002&5.29E-4&0.035\\
 \tableline
G/S&r &{\bf 0.26}&0.16&0.03 &0.07&- 0.11& 0.06\\
&P &{\bf 3.92E-6}&0.004&0.646&0.213&0.048&0.312\\
 \tableline
(1) G/S&r &- 0.09&- 0.17&- 0.09&- 0.26&0.003&0.14  \\
&P &0.243&0.036&0.277&0.001&0.965&0.076 \\
 \tableline
(2) G/S&r &{\bf 0.44}& 0.28& 0.01& 0.15& - 0.13& 0.006 \\
&P &{\bf 6.15E-8}&8.26E-4& 0.867&0.07&0.113& 0.935 \\
 \tableline
\end{tabular}
\end{center}
\end{table*}

\begin{table*}
\small
\begin{center}
\caption{The relationship between log (L H$\alpha$/L H$\beta$) and the ratios of the \ion{Fe}{2} groups.\label{tbl-6}}
\vspace*{0.3cm}
\begin{tabular}{|c| c c|}
\tableline\tableline
\vspace*{-0.2cm}
\tiny{}&\tiny{} &\tiny{}\\
\vspace*{-0.2cm}
  &&log (L H$\alpha$/L H$\beta$)\\
\tiny{}&\tiny{} &\tiny{}\\

\tableline
log (F/G)&r & -0.36 \\
&P & $<0.0001$ \\
 \tableline
  log (F/S)&r &-0.34\\
&P & $<0.0001$ \\
   \tableline
 log (G/S)&r&0.004 \\
&P &  0.962 \\
\tableline
\end{tabular}
\end{center}
\end{table*}

\begin{table*}
\small
\begin{center}
\caption{Relationships between the \ion{Fe}{2} width and the widths of the NLR, ILR, and VBLR components of the H$\alpha$ and H$\beta$ (left), and the same but for the shifts (right). Relationships are fitted with function $Y=A+B\cdot X$. The coefficient of correlations, $r$, the corresponding measure of the significance of correlations, $P$, as well as the $A$ and $B$ coefficients, are shown in the table.\label{tbl-7}}
\vspace*{0.3cm}
\begin{tabular}{|cccc|cccc|}
\tableline\tableline

 & &{\scriptsize w Fe II} &{\scriptsize Corr.}& & &{\scriptsize shift Fe II}&{\scriptsize Corr.}\\
\tableline

{\scriptsize w H$\alpha$ NLR}&{\scriptsize r}&{\scriptsize 0.01}&{\scriptsize No}&{\scriptsize sh H$\alpha$ NLR}&{\scriptsize r}&{\scriptsize 0.02}&{\scriptsize No}\\
&{\scriptsize P}&{\scriptsize 0.871}& & &{\scriptsize P}&{\scriptsize 0.84}&\\
& {\scriptsize A} &{\scriptsize  $232.71\pm27.12$  }                &    &  &{\scriptsize  A} &{\scriptsize  $149.78\pm8.88$}&\\
         &{\scriptsize B} & {\scriptsize$0.00\pm0.02$}                     &    &  &{\scriptsize B} &{\scriptsize $0.00\pm0.02$ }&\\
\tableline
{\scriptsize w H$\alpha$  ILR}&{\scriptsize r}&{\scriptsize 0.77}  &{\scriptsize Yes  }        &{\scriptsize sh H$\alpha$  ILR}&{\scriptsize r}&{\scriptsize 0.30}&{\scriptsize Yes, weak}\\
&{\scriptsize P}&{\scriptsize $<0.0001$}                               &    &    &{\scriptsize P}&{\scriptsize 0.0004}&\\
&{\scriptsize A} & {\scriptsize $-94.35\pm111.64$  }             &    &    &{\scriptsize A} &{\scriptsize $66.10\pm26.11$}&\\
&{\scriptsize B} & {\scriptsize $0.96\pm0.08$  }                 &    &    &{\scriptsize B} &{\scriptsize $0.26\pm0.07$}& \\
\tableline
{\scriptsize w H$\alpha$  VBLR}&{\scriptsize r}&{\scriptsize 0.66} &{\scriptsize Yes }         &  {\scriptsize  sh H$\alpha$  VBLR}&{\scriptsize r}&{\scriptsize 0.01}&{\scriptsize No}\\
&{\scriptsize P}&{\scriptsize $<0.0001$ }                               &   &    &{\scriptsize P}&{\scriptsize 0.930}&\\
&{\scriptsize A} & {\scriptsize $609.84\pm356.76$}                &   &    &{\scriptsize A} &{\scriptsize $-49.14\pm44.47$}&\\
&{\scriptsize B} &{\scriptsize $2.63\pm0.26$       }             &   &    &{\scriptsize B} & {\scriptsize $0.01\pm0.12$}& \\
\tableline
{\scriptsize w H$\beta$  NLR}&{\scriptsize r}&{\scriptsize 0.05 } &{\scriptsize No  }   &     {\scriptsize     sh H$\beta$  NLR}&{\scriptsize r}&{\scriptsize 0.00} &{\scriptsize No}\\
&{\scriptsize P}&{\scriptsize 0.401    }                          &   &          &{\scriptsize P}&{\scriptsize 0.928}&\\
&{\scriptsize A} & {\scriptsize $275.84\pm30.17$}           &   &         &{\scriptsize A}  & {\scriptsize $165.22\pm9.63$}&\\
&{\scriptsize B} & {\scriptsize $0.02\pm0.02$  }            &   &          &{\scriptsize B}  & {\scriptsize $0.00\pm0.03$} &\\
\tableline
{\scriptsize w H$\beta$  ILR}&{\scriptsize r}&{\scriptsize 0.73 }  &{\scriptsize Yes}          & {\scriptsize   sh H$\beta$  ILR}&{\scriptsize r}&{\scriptsize 0.39}&{\scriptsize Yes}\\
&{\scriptsize P}&{\scriptsize $<0.0001$}                                &   &    &{\scriptsize P}&{\scriptsize $<0.0001$}&\\
&{\scriptsize A} &{\scriptsize $-114.23\pm94.28$   }             &   &    &{\scriptsize A} &{\scriptsize $55.32\pm14.80$}&\\
&{\scriptsize B} &{\scriptsize $1.17\pm0.06$    }                &   &    &{\scriptsize B} &{\scriptsize $0.32\pm0.04$}& \\
\tableline                                              
{\scriptsize w H$\beta$  VBLR}&{\scriptsize r}&{\scriptsize 0.45 } &{\scriptsize Yes}  &     {\scriptsize       sh H$\beta$  VBLR}& {\scriptsize r}&{\scriptsize -0.17}& {\scriptsize No}\\
&{\scriptsize P}&{\scriptsize $<0.0001$ }                        &  &            &{\scriptsize P}&{\scriptsize 0.003}&\\
&{\scriptsize A} &{\scriptsize $2241.03\pm251.81$  }      &  &            &{\scriptsize A} & {\scriptsize $1301.87\pm96.29$}&\\
&{\scriptsize B} &{\scriptsize $1.48\pm0.17$   }         &  &            &{\scriptsize B} & {\scriptsize $-0.83\pm0.28$} &\\

\tableline
\end{tabular}
\end{center}
\end{table*}

\begin{table*}
\small
\begin{center}
\caption{Correlations between luminosities ($L$) of the \ion{Fe}{2} lines (total \ion{Fe}{2}, $F$, $S$ and $G$ line groups) and luminosities of [\ion{N}{2}], [\ion{O}{3}] and H$\alpha$ and H$\beta$ components. Relationships are fitted with function $Y=A+B\cdot X$. The coefficient of correlations, $r$, and the corresponding measure of the significance of correlations, $P$ are shown in the table.\label{tbl-8}}
\vspace*{0.3cm}
\begin{tabular}{c c c c c c c}
\tableline\tableline
\vspace*{-0.2cm}
\tiny{}&\tiny{} &\tiny{}&\tiny{}&\tiny{} &\tiny{}&\tiny{}\\
\vspace*{-0.2cm}
 & & log(L Fe II total) & log(L F) & log(L S) & log(L G) &log(L I Zw 1 group)\\
\tiny{}&\tiny{} &\tiny{}&\tiny{}&\tiny{} &\tiny{}&\tiny{}\\
\tableline

 log(L [N II]) &r &0.73 &  0.72&0.67 &  0.73&  0.72\\
   &P& $<0.0001$ & $<0.0001$& $<0.0001$ & $<0.0001$ & $<0.0001$ \\
\tableline
log(L [N II]/L [O III]) &r&  0.06 & 0.08&  0.08 & 0.10& 0.02\\
   & P& 0.51 &0.38& 0.35 & 0.25 & 0.84 \\
\tableline
log(L H$\alpha$ NLR)&r &  0.76&  0.74&  0.71&  0.75&  0.77\\
      &  P&$<0.0001$ &$<0.0001$& $<0.0001$ & $<0.0001$ & $<0.0001$\\
\tableline
log(L H$\alpha$ ILR)&r & 0.91 & 0.90& 0.84 & 0.89& 0.90\\
&P&$<0.0001$ & $<0.0001$& $<0.0001$ & $<0.0001$ & $<0.0001$\\
\tableline
log(L H$\alpha$ VBLR)&r & 0.90 & 0.89& 0.82& 0.87& 0.90\\
&P&$<0.0001$ & $<0.0001$& $<0.0001$ & $<0.0001$ & $<0.0001$\\
\tableline
log(L H$\beta$ NLR)&r &  0.64&  0.63&  0.61&  0.63&  0.64\\
& P&$<0.0001$ &$<0.0001$& $<0.0001$ & $<0.0001$ & $<0.0001$\\
\tableline
log(L H$\beta$ ILR)&r & 0.93 & 0.93& 0.88& 0.92& 0.92\\
&P&$<0.0001$ & $<0.0001$& $<0.0001$ & $<0.0001$ & $<0.0001$\\
\tableline
log(L H$\beta$ VBLR)&r & 0.95 & 0.95& 0.90& 0.93& 0.95\\
&P&$<0.0001$ & $<0.0001$& $<0.0001$ & $<0.0001$ & $<0.0001$\\
\tableline
\end{tabular}
\end{center}
\end{table*}

\begin{table*}
\small
\begin{center}

\caption{The same as in Table 8, but for EWs.\label{tbl-9}}
\vspace*{0.3cm}
\begin{tabular}{c c c c c c c}
\tableline\tableline
\vspace*{-0.2cm}
\tiny{}&\tiny{} &\tiny{}&\tiny{}&\tiny{} &\tiny{}&\tiny{}\\
\vspace*{-0.2cm}
 & & EW Fe II total & EW F & EW S & EW G &EW I Zw 1 group \\
\tiny{}&\tiny{} &\tiny{}&\tiny{}&\tiny{} &\tiny{}&\tiny{}\\
\tableline
 EW [N II]  &r &0.03 &-0.00&0.05 &  0.07&  -0.01\\
   &P& 0.74 & 0.98&0.58 & 0.40 & 0.89 \\
\tableline
EW [O III] &r&  -0.39 & -0.40&  -0.37 & -0.42 & -0.20\\
   & P& $<0.0001$ &$<0.0001$& $<0.0001$ & $<0.0001$ & 0.0006 \\
\tableline
EW [O III]/EW H$\beta$&r&  -0.46 & -0.47&  -0.43 & -0.45 & -0.28\\
   & P& $<0.0001$ &$<0.0001$& $<0.0001$ & $<0.0001$ & $<0.0001$ \\
\tableline
EW H$\alpha$ NLR&r &  -0.02&  -0.06& -0.01&  0.01&  0.00\\
      &  P&0.84 &0.49& 0.90 & 0.89 & 0.99\\
\tableline
EW H$\alpha$ ILR&r & -0.15 & -0.12& -0.12 & -0.18& -0.13\\
&P& 0.08& 0.15 & 0.15 & 0.04& 0.14\\
\tableline
EW H$\alpha$ VBLR&r & -0.17 & -0.19& -0.18& -0.19& -0.03\\
&P&0.05 & 0.03& 0.03 &0.03  & 0.69\\
\tableline
EW H$\beta$ NLR&r &  0.01&  0.01&  -0.09&  -0.04&  0.15\\
& P&0.90 &0.91& 0.11 & 0.46 & 0.01\\
\tableline
EW H$\beta$ ILR&r & -0.02 & 0.02& 0.07& -0.07& -0.08\\
&P&0.70 & 0.69& 0.25 & 0.24 & 0.18\\
\tableline
EW H$\beta$ VBLR&r & 0.12 & 0.11& 0.11& 0.04& 0.21\\
&P&0.03 & 0.07& 0.07 & 0.52 & 0.0003\\
\tableline
\end{tabular}
\end{center}
\end{table*}

\begin{table*}
\small
\begin{center}
\caption{The equivalent width of \ion{Fe}{2} vs. the widths of the H$\beta$ components.\label{tbl-10}}
\vspace*{0.3cm}
\begin{tabular}{|c|c c|}

\tableline\tableline
\vspace*{-0.2cm}
\tiny{}&\tiny{} &\tiny{}\\
\vspace*{-0.2cm}
  &&EW \ion{Fe}{2}\\
\tiny{}&\tiny{} &\tiny{}\\
\tableline
width H$\beta$ NLR&r & 0.30 \\
&P &$<0.0001$  \\
\tableline
width H$\beta$ ILR&r &-0.31\\
&P &$<0.0001$ \\
\tableline
width H$\beta$ VBLR &r&-0.34 \\
&P &  $<0.0001$ \\
\tableline
H$\beta$ FWHM&r&-0.30 \\
&P &  $<0.0001$ \\
\tableline
\end{tabular}
\end{center}
\end{table*}

\begin{deluxetable}{cccccccc}
\tabletypesize{\footnotesize}
\tablecaption{Correlations for the total sample (first and second column), the sub-sample with FWHM $H\beta<$3000 $\mathrm{kms^{-1}}$ (1), and sub-sample with FWHM $H\beta>$3000 $\mathrm{kms^{-1}}$ (2).  The correlations with P$<$0.0001 are given in bold print. \label{tbl-11}}

\tablehead{
\colhead{} & \colhead{} & \colhead{log $\lambda L_{5100}$} & \colhead{log z} &\colhead{(1) log $\lambda L_{5100}$}& \colhead{(1) log z} &\colhead{(2) log $\lambda L_{5100}$}&\colhead{(2) log z}
 }
\startdata
log (EW [O III]/EW Fe II)&r &{\bf -0.46}&{\bf -0.48}&{\bf -0.44}&{\bf -0.45} &{\bf -0.55}&{\bf -0.56}  \\
&P &{\bf 0}&{\bf 0 }&{\bf 5.67E-9}&{\bf 1.89E-9}&{\bf 1.69E-12}&{\bf 6.40E-13}  \\
\tableline
log (EW [O III]/EW H$\beta$)&r &{\bf-0.49}&{\bf-0.48}&{\bf-0.43}&{\bf-0.43}&{\bf-0.50}&{\bf-0.49 } \\
&P &{\bf 0}&{\bf 0}&{\bf 1.66E-8}&{\bf 2.00E-8}&{\bf 2.14E-10}&{\bf 5.85E-10 } \\
\tableline
log (EW H$\beta_{\mathrm{total}}$/EW Fe II)&r &-0.19&{\bf-0.28}&-0.28 &{\bf-0.31} &{\bf-0.45}&{\bf-0.49} \\
&P &0.001&{\bf 1.30E-6}&4.00E-4&{\bf 5.47E-5}&{\bf 2.40E-8}&{\bf 9.53E-10 }\\
\tableline
log (EW H$\beta$ NLR/EW Fe II)&r &{\bf-0.42} &{\bf-0.42} &{\bf-0.52}&{\bf-0.51}&-0.31&-0.31  \\
&P &{\bf 2.35E-14}&{\bf 4.75E-14}&{\bf 2.42E-12}&{\bf 7.17E-12} &2.22E-4&1.97E-4 \\
\tableline
log (EW [O III])&r &{\bf-0.43}&{\bf-0.42} &{\bf-0.33} &{\bf-0.32} &{\bf-0.54}&{\bf-0.53} \\
&P &{\bf4.00E-15}&{\bf 1.60E-14}&{\bf1.89E-5}&{\bf4.16E-5}&{\bf3.35E-12}&{\bf1.37E-11 } \\
\tableline
log (EW $\mathrm{Fe II_{total}}$)&r & {\bf 0.30}&{\bf 0.39}& {\bf 0.47}&{\bf 0.53}& {\bf 0.36}&{\bf 0.41}  \\
&P &{\bf 1.89E-7}&{\bf 3.82E-12}&{\bf 7.05E-10}&{\bf 6.09E-13}&{\bf 1.06E-5}&{\bf 4.16E-7} \\
\tableline
log (EW Fe II F)&r  &0.16&{\bf 0.26} &0.23&0.28 &0.27&{\bf 0.35}\\
&P &0.004&{\bf 5.99E-6}&0.003&2.81E-4&0.001&{\bf 2.36E-5}\\
\tableline
 log (EW Fe II S)&r&{\bf 0.33}&{\bf 0.40}&{\bf 0.52}&{\bf 0.58}&{\bf0.37}&{\bf 0.39} \\
&P & {\bf 3.87E-9} & {\bf 1.04E-12 }&{\bf 1.87E-12} &{\bf 8.88E-16}&{\bf 4.57E-6} &{\bf 1.37E-6} \\
\tableline
log (EW Fe II G)&r&{\bf 0.44}&{\bf0.50}&{\bf0.61}&{\bf 0.65}&{\bf 0.50}&{\bf 0.51}\\
&P &{\bf 1.78E-15}&{\bf 0}&{\bf 0}&{\bf 0 }&{\bf 2.52E-10}&{\bf 9.89E-11}\\
\tableline
log (EW Fe II I Zw 1 group)&r&{\bf 0.32}&{\bf 0.23}&{\bf 0.31}&{\bf 0.39}&0.13&0.18\\
&P &{\bf 1.77E-8}&{\bf 4.58E-5}&{\bf 8.33E-5}&{\bf 4.29E-7}& 0.120&0.031\\
\tableline
log (EW H$\beta$ NLR)&r&{\bf-0.36}&{\bf-0.33}&{\bf-0.44}&{\bf-0.41}&-0.21&-0.20\\
&P &{\bf 2.05E-10}&{\bf 6.78E-9}&{\bf 8.87E-9}&{\bf 8.79E-8}&0.010&0.015\\
\tableline
log (EW H$\beta$ ILR) &r&0.08&0.03&-0.04&-0.04&-0.11&-0.08\\
&P &0.161&0.545&0.611&0.603&0.21&0.315\\
\tableline
log (EW H$\beta$ VBLR) &r&0.16&0.17&{\bf 0.41}&{\bf 0.44}&-0.09&-0.09\\
&P &0.006&0.002&{\bf 6.98E-8}&{\bf 9.53E-9}&0.292&0.271\\
\tableline
\enddata
\end{deluxetable}

\end{document}